\documentclass[12pt]{iopart}
\usepackage[a4paper]{geometry}
\usepackage{graphicx,amsmath,bm}
\usepackage{amsfonts}
\usepackage{amssymb}
\usepackage{subcaption}
\usepackage{cite}

\usepackage{color}

\usepackage[hidelinks]{hyperref}

\newcommand{\av}[1]{\langle #1 \rangle}
  
\begin{document}

\title[Concatenated stochastic heat engines]{Exploring outputs from concatenated stochastic heat engines}

\author{Aradhana Kumari$^1$, Rahul Marathe$^2$ and Sourabh Lahiri$^1$}

\address{$^1$Department of Physics, Birla Institute of Technology Mesra, Jharkhand 835215, India}
\address{$^2$Department of Physics, Indian Institute of Technology Delhi, Hauz Khas 110 016, New Delhi, India}

\eads{\mailto{aradhanakumari2546@gmail.com}, \mailto{maratherahul@physics.iitd.ac.in}, \mailto{sourabhlahiri@bitmesra.ac.in}}

\vspace{10pt}
\begin{indented}
\item[]
\end{indented}

\begin{abstract}
Recent works on the concatenation of two simple heat engines have shown that it may lead to non-monotonic variations in the efficiency and power with parameters like driving amplitudes and asymmetries in cycle periods. Motivated by this study, we investigate the effect of the concatenation between two stochastic heat engines where colloidal particles have been trapped in harmonic potentials. The stiffness parameters of each engine are varied cyclically, but with different cycle periods, with a common thermal bath that acts as a sink for the first engine but as a source for the second. 
    We consider two types of protocols, first where the trap strength undergoes sudden jumps, and the second where it varies linearly with time.
    In both we find several non-trivial effects, like the the non-monotonic functional dependence of the engine outputs on several parameters used in the setup. For a protocol that varies linearly with time, the concatenation leads to enhanced output power as compared to a single effective engine, in a suitable range of parameters. It has been shown that the output from the combined system shows a peak with respect to the asymmetry in cycle times of the engines that have been concatenated. A general relation of the efficiency of an arbitrary number of concatenated  engines driven quasistatically has been provided.
\end{abstract}


\submitto{Journal of Statistical Mechanics: Theory and Experiment}

 \section{Introduction}
 
 The challenge of theoretical and experimental study on heat engines, classical and quantum, and their potential benefits have attracted a lot of attention over the past decade or so \cite{sei08_epl,Rana2014,Abah2014,Verley2014,bechinger2012,Koski2014,Abah2016,Goold2019}. The single-particle manipulation techniques have greatly boosted the experimental study of their thermodynamic properties \cite{Ritort2005,Ritort2006}. The important difference from macroscopic thermal engines is the presence of appreciable thermal noise at microscopic scales, thus implying that microscopic thermal machines are not just the naive scaled-down version of their macroscopic counterparts \cite{Jarzynski2011}.

 The working system in most of the theoretical investigations on classical heat engines and refrigerators includes a colloidal particle following an overdamped Langevin equation. Several generalizations have been made thereafter, including the case of replacing the simple Brownian particles with self-propelled particles of various types, like the active Brownian particles (ABPs) \cite{Saha_2018, Saha_2019, Holubec_2020, Majumdar_2022, Marathe_2023}, active Ornstein-Uhlenbeck particles (AOUPs) \cite{lahiri2020}, and run-and-tumble particles (RTPs) \cite{Marathe_2023, Aradhana_2021}. Several extensions have been made in the quantum regime, where phenomena like squeezing \cite{Klaers2017,wang_2019,Kumar2022}, entanglement \cite{zhang_2007,brunner2014} and coherence \cite{Shi_2020,Scully2001,Scully2011,scully2003extracting} have been studied.

 The optimization of such thermal machines has also received a major attention \cite{salas_2010,Holubec_2014,puglisi_2021}, since it helps in reducing the dissipation during nonequilibrium protocols, leading to increased practical applications of such machines. The optimal protocols (externally controlled time-variation of a single or more parameters) for a harmonic potential \cite{sei08_epl}  and for log-harmonic potential \cite{Holubec_2014}  have been derived. The efficiency at maximum power has been studied in several works \cite{Van_den_Broeck_2005,sei08_epl,Esposito_2009,Esposito_2010}.

 In some recent works \cite{Noa_2020}, the authors focus on the dependence of efficiency and power on the driving forces acting on engines that are sequentially connected to different thermal reservoirs, for both constant and linear drives. In an extension to this work \cite{Noa_2021}, the authors introduce an asymmetry between the contact times with the different heat baths and in the magnitude of the driving forces, as possible parameters that can be used to optimize the performance. 
 Such a study is useful in the investigation of collisional engines \cite{Stable2020,Pezzutto2016}, where the working system interacts at a time with only a portion of its environment.

 In the present work, we build on this idea and construct a concatenated  engine that is placed in a harmonic trapping potential whose stiffness parameter is made time-dependent. This is in alignment with the frequently used prototypes for stochastic heat engines used in the literature \cite{sei08_epl,Rana2014,roldan2016,bechinger_2012,ajay2016_nature}. We find that as long as the engine remains in the underdamped regime, in several cases we obtain interesting features in the engine outputs, as a function of various parameters that can directly affect the performance of such engines. Unfortunately, these features disappear in the overdamped limit, thus leaving the study difficult to investigate analytically, except in a special case referred to as a \emph{jump protocol} (see Sec. \ref{sec:ProtocolTypes}).

The paper is organized as follows. In Sec. \ref{sec:Model}, we discuss the model that we are investigating by means of analytics and simulations. This includes the equations of motion governing the dynamics of the working system, and the types of protocols used. In Sec. \ref{sec:Results}, we discuss the results of two concatenated  engine subjected to the jump protocol as well as to the linear protocol. In the former case, we compare the outcomes of simulations with analytics, and the latter protocol is studied entirely using simulations. In Sec. \ref{sec:ThreeEngines}, we briefly discuss the concatenation of three engines. Sec. \ref{sec:recursion} discusses the generalizations to the case where higher number of engines are concatenated . Finally, we conclude in Sec. \ref{sec:Conclusions}.

 \section{Model}
 \label{sec:Model}

 \subsection{Equations of motion}
 
\begin{figure}[h!]
 \centering
 \includegraphics[width=1\textwidth,height=8cm]{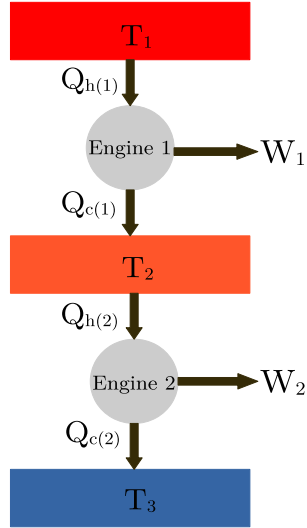}
 \caption{Schematic diagram showing two concatenated  stochastic heat engines}
 \label{fig:two_coupled_engine}
\end{figure}

 The schematic diagram of the concatenated  engines has been provided in Fig. \ref{fig:two_coupled_engine}. 
 The first engine (engine 1) operates between two thermal baths at temperature $T_1$ (hot bath) and $T_2$ (cold bath). The cycle time for the first engine continues from $t=0$ to $t=\tau_1$. The expansion and compression processes are each equal to $t=\tau_1/2$. 
  The second engine (engine 2) runs between baths at temperatures $T_2$ and $T_3$, where $T_2>T_3$, and its cycle time is of duration $\tau_2$ and runs from $t=\tau_1$ to $t=\tau_1+\tau_2$.
  Again, the cycle time of $\tau_2$ is equally divided among the expansion step and the compression step.
  The time-dependence of the stiffness parameter for expansion and compression steps will be denoted by the functions $k_{e,i}(t)$ and $k_{c,i}(t)$,  where $i=1,2$ denotes parameters for the first and second engines respectively.

The interaction between the engines comes through the fact that the steady-state distribution reached at the end of $\tau_1$ is different from the case where engine 2 runs independently between the same thermal reservoirs (i.e., between temperatures $T_2$ and $T_3$). The same statement holds true at the end of the full cycle of the combined engines, i.e. at time $\tau_1+\tau_2$, where the steady-state distribution is not the same as the initial distribution for engine 1 running independently. In this way, the presence of each of the engines in the concatenated system affects the other engine.

 We denote the position and velocity of the particle in the first engine by $x_1$ and $v_1$ respectively, while that for the second engine by $x_2$ and $v_2$ respectively. The equations of motion for the first engine are 
\begin{alignat}{2}
 \dfrac{dx_1}{dt} &=v_1;\hspace{20pt} \dfrac{dv_1}{dt} =\dfrac{1}{m}[-\gamma v_1-k_{e,1}(t)x_1+\xi_1(t)], \hspace{40pt}  && t \in (0,\tau_1/2); \nonumber\\
 \dfrac{dx_1}{dt} &=v_1;\hspace{20pt} \dfrac{dv_1}{dt} =\dfrac{1}{m}[-\gamma v_1-k_{c,1}(t)x_1+\xi_2(t)],  && t \in (\tau_1/2,\tau_1).
 \label{eq:FirstEngine}
\end{alignat}
and for the second engine are
\begin{alignat}{2}
 \dfrac{dx_2}{dt} &=v_2;\hspace{20pt} \dfrac{dv_2}{dt}=\dfrac{1}{m}[-\gamma v_2-k_{e,2}(t)x_2+\xi_2(t)], \hspace{40pt} && t \in \big(\tau_1,\tau_1+\tau_2/2\big); \nonumber\\
 \dfrac{dx_2}{dt} &=v_2;\hspace{20pt} \dfrac{dv_2}{dt}=\dfrac{1}{m}[-\gamma v_2-k_{c,2}(t)x_2+\xi_3(t)], && t \in \big(\tau_1+\tau_2/2,\tau_1+\tau_2\big).
 \label{eq:SecondEngine}
\end{alignat}
The Gaussian white noise terms follow the relations $\langle\xi_i(t)\rangle=0$ and $\langle\xi_i(t)\xi_j(t')\rangle= 2\gamma T_i\delta_{ij}\delta(t-t')$, where $i,j=1,2,3$. 
For convenience, the Boltzmann constant has been set to unity throughout this article.

\subsection{Protocols used} \label{sec:ProtocolTypes}
We consider two simple functional forms of the time dependent stiffness parameter. These are the popular functional forms among the ones used in the literature \cite{lahiri2020,Aradhana_2021,Saha_2019}.
\begin{enumerate}
    \item \textbf{Sudden jump:} 
    The stiffness parameter undergoes a sudden jump mid-way during the expansion and compression processes. For engine 1, this can be mathematically stated as: 
    \begin{alignat}{2}
        k_{e,1} &= k_0, \hspace{1cm} && t\le \tau_1/4; \nonumber\\
        k_{e,1} &= k_0/2, && \tau_1/4 < t \le \tau_1/2; \nonumber\\
        k_{c,1} &= k_0/2, && \tau_1/2 < t \le 3\tau_1/4; \nonumber\\
        k_{c,1} &= k_0, &&  3\tau_1/4 < t \le \tau_1.
        \label{eq:Jump}
    \end{alignat}
    Similar equations can be set up for the second engine, where the jumps take place at time instances $t=\tau_1+\tau_2/4$ (expansion) and $t=\tau_1+3\tau_2/4$ (compression), respectively. The form of stiffness parameter for the complete cycle is shown in Fig. \ref{fig:stiff_jump}. 
   The protocol has been sketched with the colors red, orange and blue, in order to indicate respectively the temperatures $T_1$, $T_2$ and $T_3$, at which the process is taking place.
    This model is analytically solvable and hence can be used to verify the agreement with our simulations.

     \begin{figure}[h!]
 \centering
 \includegraphics[width=0.55\textwidth]{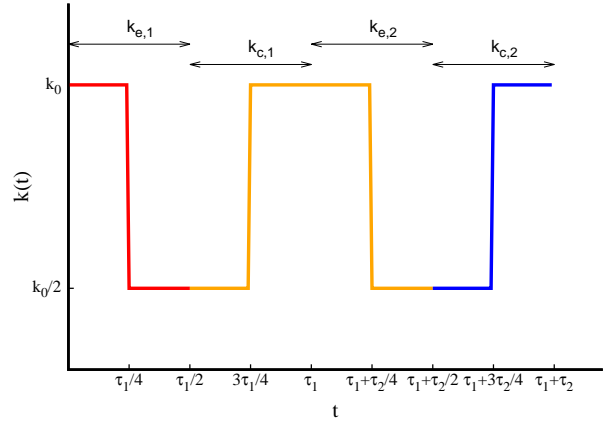}
 \caption{Variation in the stiffness parameter with time for sudden jump process. The colors red, orange and blue indicate the temperatures $T_1$, $T_2$ and $T_3$, respectively.}
 \label{fig:stiff_jump}
\end{figure}
 \paragraph{}

    \item \textbf{Linear variation:} 
    In this case, the stiffness parameter decreases linearly with time during the expansion half for each engine, and increases during the other half. For the first engine, the relations become
    \begin{alignat}{2}
        k_{e,1}(t) &= k_0(1-t/\tau_1), \hspace{2cm} && t\le \tau_1/2; \nonumber\\
        k_{c,1}(t) &= k_0(t/\tau_1), && \tau_1/2 < t \le \tau_1; \nonumber\\
        k_{e,2}(t) &= k_0[1-(t-\tau_1)/\tau_2], && \tau_1< t \le \tau_1+\tau_2/2; \nonumber\\
        k_{c,2}(t) &= k_0[(t-\tau_1)/\tau_2], && \tau_1+\tau_2/2< t \le \tau_1+\tau_2.
        \label{eq:Ramp}
    \end{alignat}
    Here, $k_{e,i}$ and $k_{c,i}$ (with $i=1,2$) are the stiffness parameters of the $i^{\rm th}$ engine during the expansion and compression strokes, respectively. The time-variation of this protocol is shown in Fig. \ref{fig:stiff_linear}. 
    As in Fig. \ref{fig:stiff_jump}, the colors indicate the corresponding temperatures at which the process takes place. The system is not analytically solvable in this case, and we need to investigate it by means of simulations.

      \begin{figure}[!ht]
 \centering
 \includegraphics[width=0.55\textwidth]{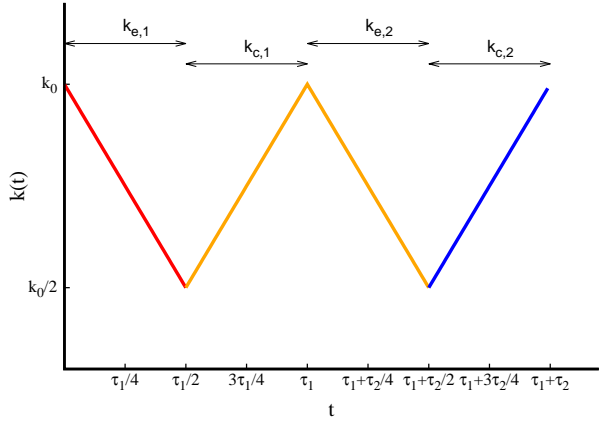}
 \caption{Variation in linearly dependent stiffness parameter with time. The colors red, orange and blue indicate the temperatures $T_1$, $T_2$ and $T_3$, respectively.}
 \label{fig:stiff_linear}
\end{figure}
 
\end{enumerate}

\subsection{Work and Heat}

The average input and output work are defined as per the prescription of stochastic energetics provided in \cite{sek98,sekimoto}. It is to be noted that for our system of concatenated  engines, the  \emph{input} work is equal to the output work obtained from the first engine, since this work is fed into the input of the second engine \cite{Noa_2020}. The output work, on the other hand, is the one that is extracted from the second engine, since that is the overall output of the concatenated  system. We define $\sigma_e(t)$ and $\sigma_c(t)$ to be the variances $\langle x^2(t)\rangle$ (angular brackets denote ensemble averages) during the expansion and compression steps, respectively. The mean input and output works are as follows:
\begin{align}
 \av{W_{1}} &= \dfrac{1}{2}\int_{0}^{\tau_1/2}\dot{k}_{e,1}(t)\sigma_e(t) dt+\dfrac{1}{2}\int_{\tau_1/2}^{\tau_1}\dot{k}_{c,1}(t)\sigma_c(t) dt \nonumber\\
 \av{W_{2}} &= \dfrac{1}{2}\int_{\tau_1}^{\tau_1+\tau_2/2}\dot{k}_{e,2}(t)\sigma_e(t) dt+\dfrac{1}{2}\int_{\tau_1+\tau_2/2}^{\tau_1+\tau_2}\dot{k}_{c,2}(t)\sigma_c(t) dt.
\end{align}
The magnitudes of absorbed heats (note that the actual values of the absorbed heats are obtained by adding a minus sign, as per our conventions) of the first and second engines into the hotter baths, denoted respectively by $Q_{h(1)}$ and $Q_{h(2)}$, are readily obtained by applying the first law:
\begin{align}
 \av {Q_{h(1)}}&=\av{W_{h(1)}}-\av {\Delta E_{h(1)}}\nonumber\\
 \av {Q_{h(2)}}&=\av{W_{h(2)}}-\av {\Delta E_{h(2)}},
\end{align}
where $W_{h(1)}$ and $W_{h(2)}$ are the work done during the expansion processes of engines 1 and 2, respectively. Similar notations have been used for the changes in internal energy as well.
The change in internal energy is defined as the difference between the total (potential plus kinetic) energies at the beginning and at the end of the expansion step:
\begin{align}
 \av {\Delta E_{h(1)}}=&\bigg[\frac{1}{2}k(\tau_1/2)\sigma(\tau_1/2)+\dfrac{1}{2} m \sigma_v(\tau_1/2)\bigg]-\bigg[\frac{1}{2}k(0)\sigma(0)+\frac{1}{2} m \sigma_v(0)\bigg] \nonumber\\
 \av{\Delta E_{h(2)}}=&\bigg[\dfrac{1}{2}k(\tau_1+\tau_2)\sigma(\tau_1+\tau_2)+\dfrac{1}{2}m \sigma_v(\tau_1+\tau_2)\bigg]\nonumber\\
 &-\bigg[\dfrac{1}{2}k(\tau_1+\tau_2/2)\sigma(\tau_1+\tau_2/2)+\dfrac{1}{2}m \sigma_v(\tau_1+\tau_2/2)\bigg].
\end{align}

The efficiency of concatenated  heat engine is defined as:
\begin{align}
 \eta=\dfrac{-\av {W_{2}}}{\av {W_{1}}-\av{Q_{h(1)}}-\av{Q_{h(2)}}} = \frac{W_{\rm out}}{-W_{\rm in}+Q_{\rm in} + Q_{\rm out}},
 \label{eq:eta}
\end{align}
where we have defined $W_{\rm in}\equiv -\av{W_1}$, $W_{\rm out}\equiv -\av{W_2}$, $Q_{\rm in} \equiv -\av{Q_{h(1)}}$, and $Q_{\rm out}\equiv -\av{Q_{h(2)}}$. The minus sign appears due to our convention to consider the heat dissipated by the system and the work done on the system as positive.

\section{Results and Discussions} \label{sec:Results}

\subsection{Case 1: Sudden jump protocol} 

We first choose case 1 outlined in Sec. \ref{sec:Model}, where the expansion and compression halves involve sudden jumps in the stiffness parameters at the mid-point of the respective halves. 
The concatenated engine subjected to the jump protocol, being exactly solvable, can be used to benchmark our codes. Once the accuracy of our simulations is established, we use the code to investigate the less trivial cases where the system has reached time-periodic its steady state, both for the jump as well as the linear protocols.

The Langevin equations can be conveniently written as the following matrix equation:\\
\begin{align}
    \frac{d\textbf{A}}{dt} &= -\textbf{M}\cdot \textbf{A} + \bm{\Xi(t)},
    \label{eq:ExpansionAnalytical}
\end{align}
For instance, for the expansion step of the first engine, the matrices are
\begin{align}
    \textbf{A} = 
    \begin{pmatrix}
    x \\ v
    \end{pmatrix};
    \hspace{0.5cm}
    \textbf{M} =
    \begin{pmatrix}
    0 & -1\\
    k_{e,1}/m & \gamma/m
    \end{pmatrix};
    \hspace{0.5cm}
    \bm{\Xi(t)} = 
    \begin{pmatrix}
    0 \\ \xi_1(t)/m
    \end{pmatrix}.
\end{align}
Here, the values of $k_{e,1}$ are given by Eq. \eqref{eq:Jump}. The formal solution to the above equation is given by
\begin{align}
    \textbf{A}(t) &= \textbf{A}(0)e^{-\textbf{M}t'} + \int_0^t  e^{-\textbf{M}(t-t')}\bm{\Xi}(t') dt'.
    \label{eq:MatrixEquation}
\end{align}
If the initial conditions for the first cycle are given by $x(0)=0$ and $v(0)=0$, then the variances for the expansion half of this cycle can be found from the relation
\begin{align}
    \left<A(t)A^T(t)\right> &=
    \begin{pmatrix}
    \langle x_1^2\rangle & \langle x_1v_1\rangle\\
    \langle x_1v_1\rangle & \langle v_1^2\rangle
    \end{pmatrix}
    \equiv 
    \begin{pmatrix}
    \sigma_{x_1} & \sigma_{x_1v_1} \\
    \sigma_{x_1v_1} & \sigma_{v_1}
    \end{pmatrix} \nonumber\\
    &=
    \int_0^t dt' \int_0^t dt'' e^{-\textbf{M}(t-t')}\langle \bm{\Xi}(t')\bm{\Xi}^T(t'')\rangle e^{-\textbf{M}^T(t-t'')}.
\end{align}
Similar solutions can be obtained for the compression half as well, and can be extended to the second engine. It is to be noted that the initial conditions for the compression half must be equal to the final values of $x$ and $v$ reached at the end of the expansion half, i.e. $A(\frac{\tau}{2}^+) = A(\frac{\tau}{2}^-)$.

The analytical expressions for the variances in position and velocity for the expansion stroke of the first engine during the first cycle are as given below. 
\begin{align}
    \sigma_x(t) &= \frac{T_1}{2k_0\alpha^2}~e^{-(\gamma+\alpha)t/m}\left[\gamma(\gamma+\alpha) e^{2\alpha t/m} + \gamma(\gamma-\alpha) -8k_0m e^{\alpha t/m} - 2\alpha^2 
    e^{(\gamma+\alpha)t/m}\right]; \nonumber\\
    \sigma_v(t) &= \frac{T_1}{m\alpha^2}~e^{-(\gamma+\alpha)t/m}\left[8k_0m e^{\alpha t/m} + 2\alpha^2 e^{(\gamma+\alpha)t/m} + \gamma(\alpha-\gamma)e^{2\alpha t/m} - \gamma(\gamma+\alpha)\right].
    \label{eq:AnalyticalVariances}
\end{align}
Similar expressions for the variances during the other strokes can be obtained, using the final values of the previous stroke as the initial values for the next stroke.  
The comparison between numerical results and simulations for both the variance $\sigma_x$ in position and the work $\langle W\rangle$ has been provided in \ref{sec:Benchmarking}. The excellent agreement between the plots complete the benchmarking of our simulations. Henceforth, the simulations would be extended to the steady state regime, where the initial transients would be removed in order to study the behaviour of the engine when its distribution functions have become time-periodic.

\begin{figure}[h!]
    \centering
    \begin{subfigure}{0.45\textwidth}
    \includegraphics[width=\linewidth]{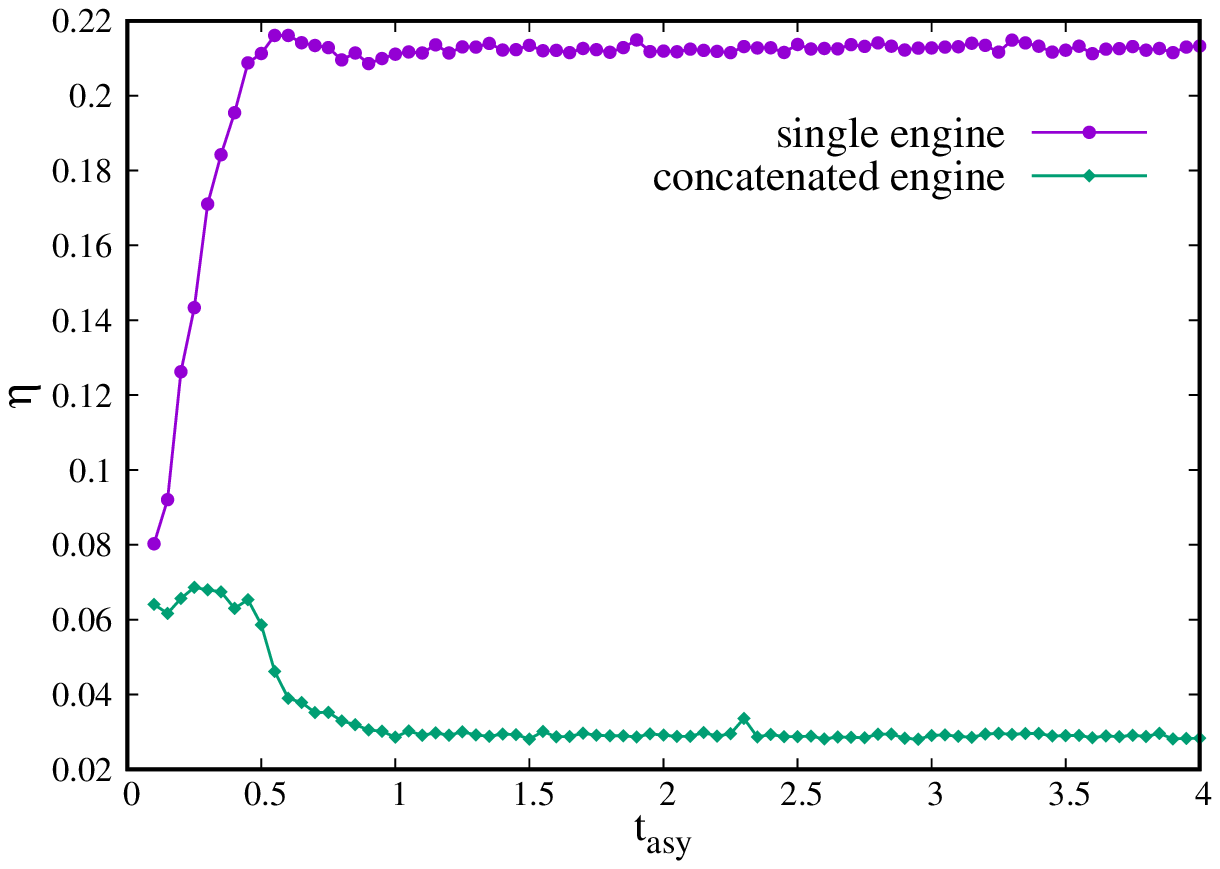}
    \caption{}
    \end{subfigure}
    \begin{subfigure}{0.45\textwidth}
    \includegraphics[width=\linewidth]{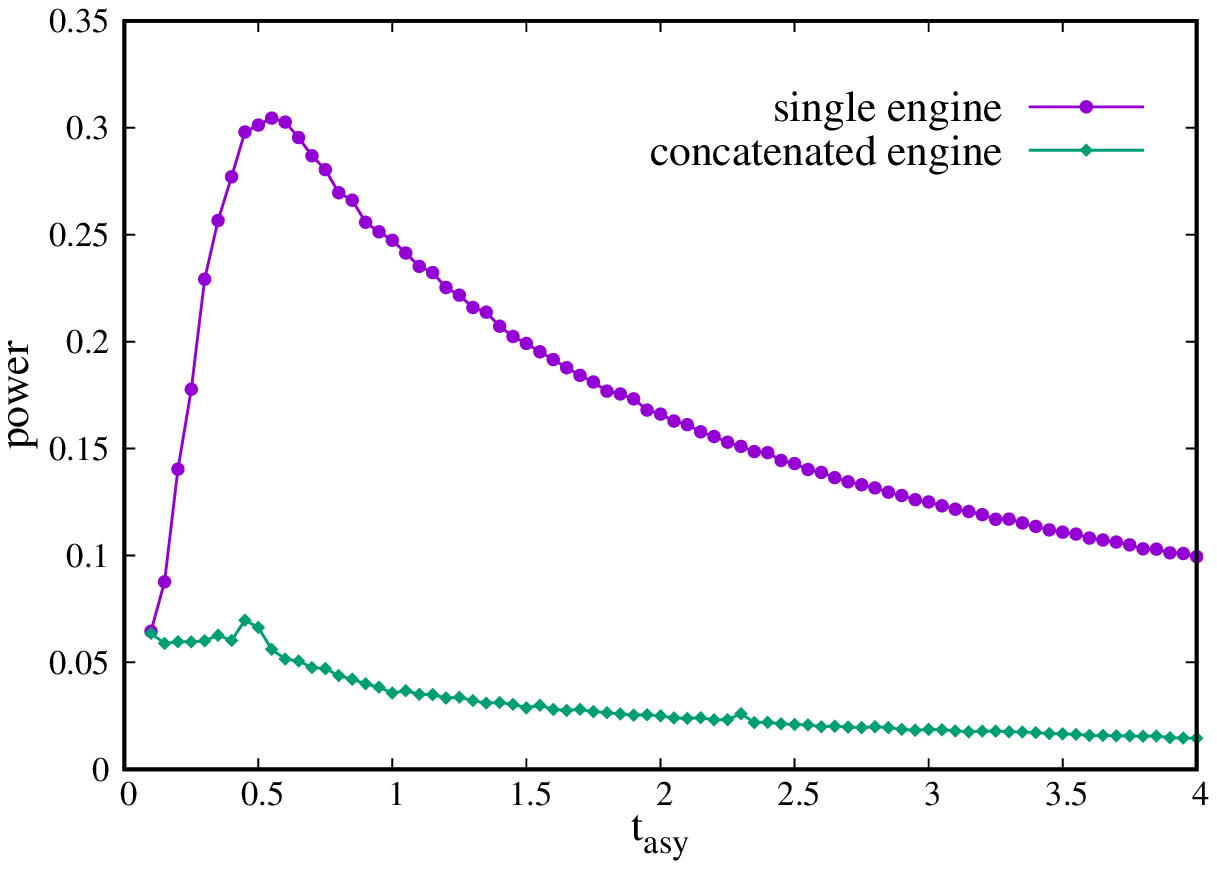}
    \caption{}
    \end{subfigure}

    \caption{(a) Plot of efficiency as a function of $t_{\rm asy}$ for sudden jump process; (b) Similar plot of power with $t_{\rm asy}$. The purple curve with solid circles correspond to the effective single engine. The parameters used are:  $k_0=10,~m=0.1,~T_1 = 2,~T_2=1,~T_3=0.1,~\tau_2=1, ~\gamma=1$.}
    \label{fig:sudden jump}
\end{figure}

The plots for efficiency and power of the concatenated  engine (in its time-periodic steady state)  operated by the jump protocol, as a function of $t_\mathrm{asy}$, has been shown in Fig. \ref{fig:sudden jump}(a) and (b) respectively, for the set of parameters mentioned in the figure caption.
We find a clear non-monotonicity in efficiency, with the maximum occurring at a small value of $t_\mathrm{asy}\ll 1$. 
For comparison, the behaviour of a single engine operating directly between the first and the third heat baths (at temperatures $T_1$ and $T_3$, respectively) has been shown (magenta line with solid circles). The single engine runs for a cycle time equal to that of the full cycle time of the concatenated  engine, i.e.
\begin{align}
    \tau_\text{single} &= \tau_1+\tau_2.
\end{align}
This means that the expansion/compression steps of engine 1 (of duration $\tau_1/2$ each) and the expansion/compression steps of engine 2 (of duration $\tau_2/2$ each) are of half the duration as the expansion/compression steps of the single engine. Note that this is the most reasonable comparison that can be made between the concatenated  and single engines in our setup. We find that the concatenated engine falls well behind the single engine when compared with respect to either efficiency or output power. We will later see that the linear protocol is a better choice for the concatenated engine (see Fig. \ref{fig:EtaPower_stiffness}).
It may further be noted that the characteristic time scale in the dynamics is given by the relaxation time of $m/\gamma$ for the velocity variable. Given that this time is $\sim 0.1$ in our simulations and $\tau_2$ has been held fixed at the value 1, the process can be assumed to be a quasistatic one when $t_{\rm asy}\gtrsim 1$, but will a highly nonequilibrium one when  $t_{\rm asy}\ll 1$.

Fig. \ref{fig:sudden jump}(b) shows the output power as a function of $t_{\rm asy}$. Although the non-monotonicity in the case of  concatenated  engines is not prominent, it can be clearly observed for the single engine. Next, we study the behaviour of the different thermodynamic variables appearing in Eq. \eqref{eq:eta}, in order to throw more light on the observed peak in the efficiency at a particular value of $t_{\rm asy}$.

\begin{figure}
    \centering
    \includegraphics[width=0.5\textwidth]{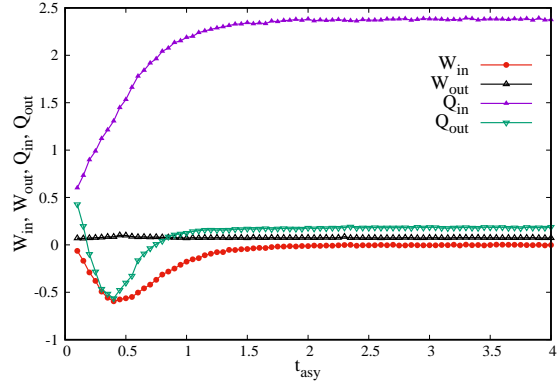}
    \caption{Plots for the different thermodynamic observables for the jump protocol, appearing in the definition of efficiency (see Eq. \eqref{eq:eta}). The parameters are the same as in Fig. \ref{fig:sudden jump}.}
    \label{fig:EfficiencyComponents}
\end{figure}

The variations with $t_{\rm asy}$ in each of the thermodynamic variables appearing in the definition of efficiency in Eq. \eqref{eq:eta} have been probed in Fig. \ref{fig:EfficiencyComponents}. We find that two of the variables that appear in the denominator of the equation, namely $W_{\rm in}$ and $ Q_{\rm out}$, show a dip at $t_{\rm asy}\approx 0.4$. Since they appear with opposite signs in the denominator of the equation, they approximately cancel each other. The peak around this region is expected on the basis of the small hump appearing in $W_{\rm out}$ and the smaller value of $Q_{\rm in}$, with the latter reaching a plateau at a higher value of  $t_{\rm asy}$.


\subsection{Case 2: Linearly varying stiffness parameters}

\begin{figure}[!ht]
    \centering
    \begin{subfigure}{0.45\textwidth}
    \includegraphics[width=\linewidth]{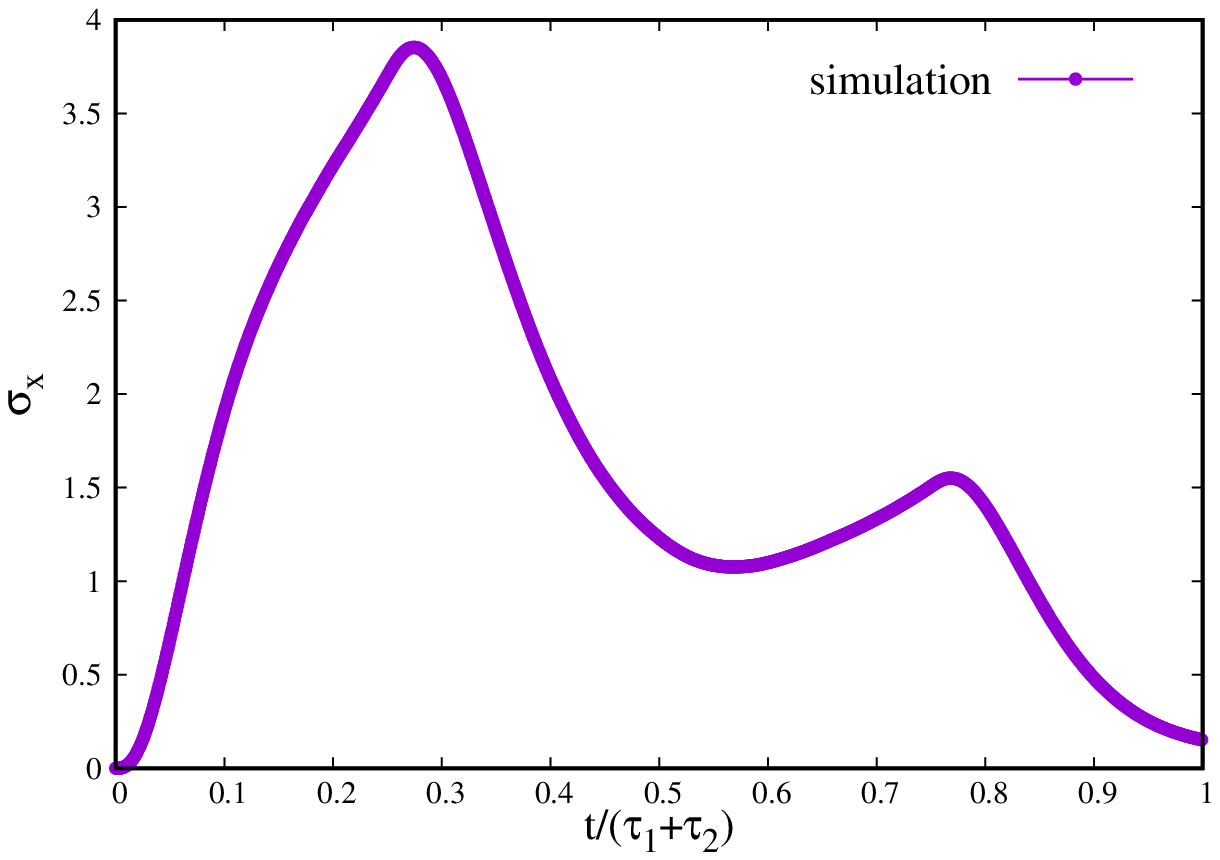}
    \caption{}
    \end{subfigure}
    \begin{subfigure}{0.45\textwidth}
    \includegraphics[width=\linewidth]{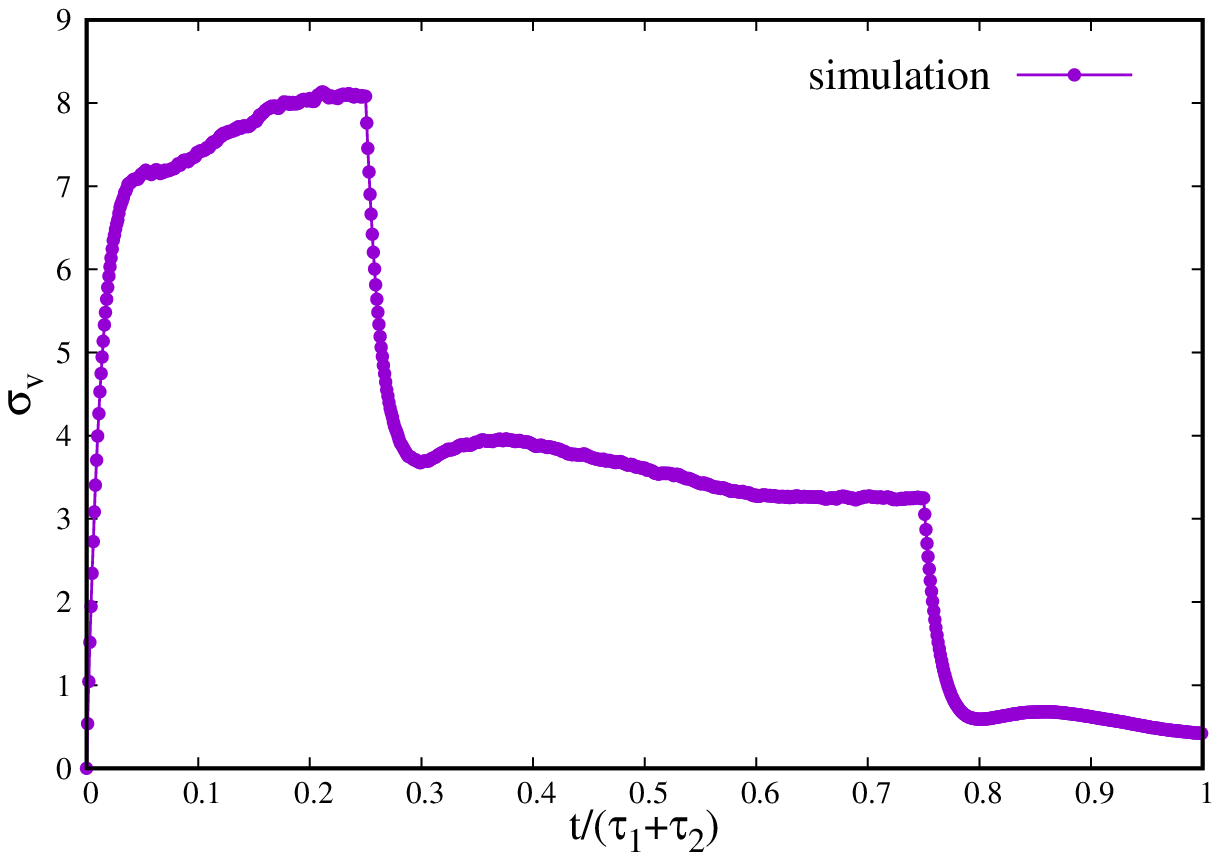}
    \caption{}
    \end{subfigure}

    \caption{(a) Plot of variance in position  for a full cycle of concatenated  engine for the case of linearly dependent stiffness parameter, when both $x$ and $v$ begin from zero at $t=0$. (b) Similar plot for variance in velocity. The parameters are: $m=0.3,~T_1 = 2.5,~T_2=1, ~T_3=0.1, ~k_0=1, ~\gamma=1, ~\tau_1=5, ~\tau_2=5$.}
    \label{fig:variance_ld}
\end{figure}
In Fig. \ref{fig:variance_ld}(a) and (b), we have shown the simulated variances in position and velocity as a function of time in the full engine cycle of duration $\tau_1+\tau_2$. A comparison with Fig. \ref{fig:Comparison} shows that the qualitative trends are similar, while are relatively close  quantitatively. Again, the two peaks in position variance correspond to the switching from expansion to compression step in the case of each of the concatenated  engines.  

\begin{figure}[h!]
    \centering
    \begin{subfigure}{0.48\textwidth}
    \includegraphics[width=\linewidth]{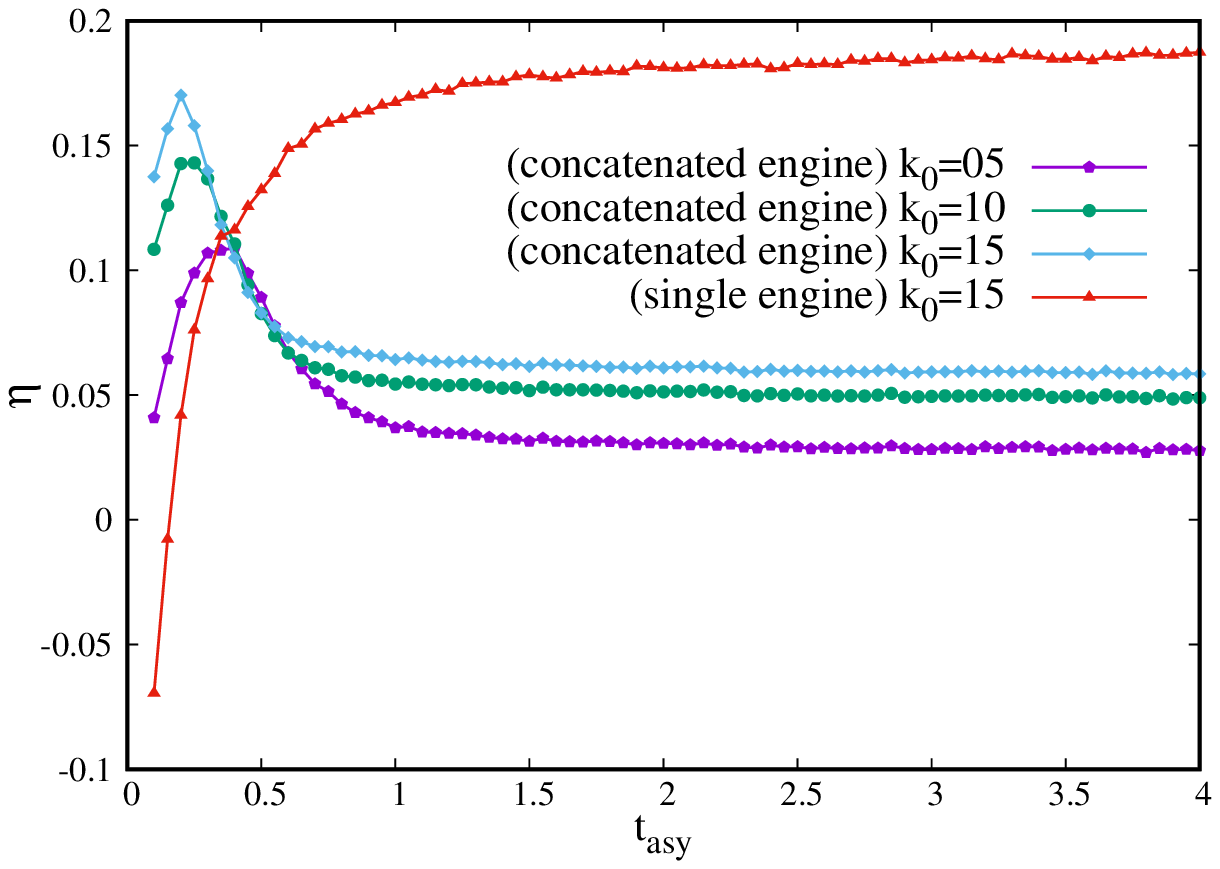}
    \caption{}
   
    \end{subfigure}
    \begin{subfigure}{0.48\textwidth}
    \includegraphics[width=\linewidth]{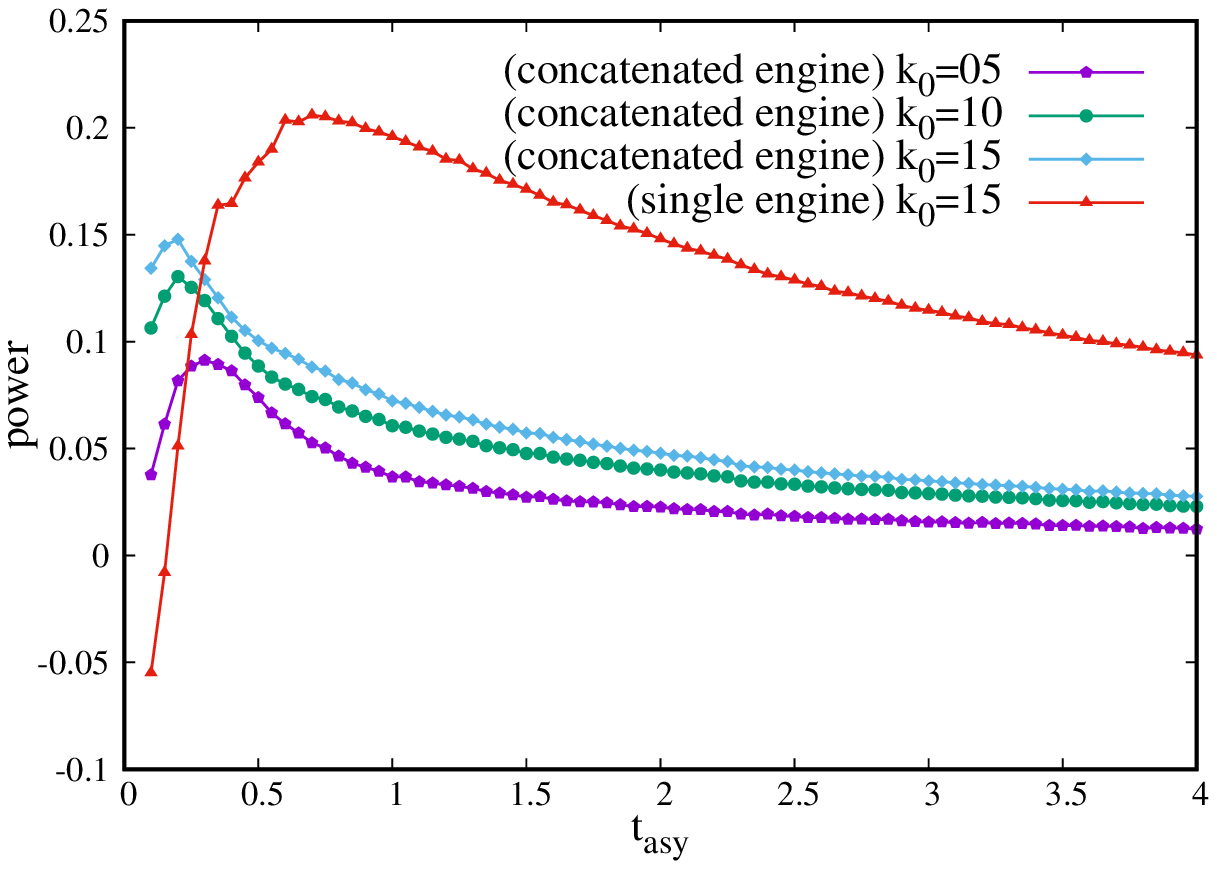}
    \caption{}
   
    \end{subfigure}
  \caption{(a) The plots show efficiency for the concatenated  engine with linearly varying protocol, with asymmetric time for $k_0=5$ (purple line), $k_0=10$ (green line), $k_0=15$ (cyan line). (b) Similar plot for the magnitude of extracted power. The red curve with triangles correspond to the effective single engine.} The other used parameters are: $T_1=2$, $T_2=1$, $T_3=0.1$, $m=0.1$.
      \label{fig:EtaPower_stiffness}
\end{figure}
The efficiency of a concatenated  heat engine with the time asymmetry $t_{\rm asy}$ is plotted in Fig. \ref{fig:EtaPower_stiffness}(a). The red solid line with solid triangles is the reference curve that shows the variation of $\eta$ with $t_{\rm asy}$ for a single engine working between $T_1$ and $T_3$ (see Fig. \ref{fig:two_coupled_engine}). 
The plot shows that for $t_{\rm asy}\gtrsim 0.5$, the efficiency of concatenated  engine is less than from that of the single engine with same parameters, as expected from equilibrium thermodynamics (see Sec. \ref{sec:recursion}). In the region where  $t_{\rm asy}< 0.5$, the plot shows non-monotonic variation of efficiency with $t_{\rm asy}$, where it outperforms the single engine. 
Note that this behaviour is in stark contrast to the case of jump protocol (see Fig. \ref{fig:sudden jump}), where the efficiency of the concatenated engine remained less than that of a single engine throughout the range of $t_{\rm asy}$.
A peak is observed for all the three values of $k_0$ but at different values of $t_{\rm asy}$, all being present in the region $\tau_1<\tau_2$. 
At the same time, $W_{\rm out}$ also increases in this region, as can be seen from
Fig. \ref{fig:EtaPower_stiffness}(b). We clearly see that the region $t_{\rm asy}<0.5$ shows a peak in the power for all values of $k_0$ used in our simulations. We further note from the figure that higher
values of $k_0$ lead to extraction of higher power from the concatenated engine. However,
except near the peaks, the single engine (red solid line with triangles) outperforms
the concatenated ones, in terms of output power. 

\begin{figure}
    \centering
    \includegraphics[width=0.5\textwidth]{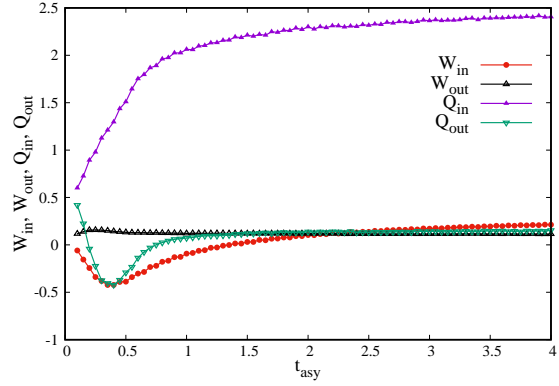}
    \caption{Plots for the different thermodynamic observables appearing in the definition of efficiency (see Eq. \eqref{eq:eta}) for the linear protocol. The parameters are the same as in Fig. \ref{fig:EtaPower_stiffness}.}
    \label{fig:components_linear}
\end{figure}

In order to understand the peaking behavior close to the left end of the plot, in Fig. \ref{fig:components_linear} we have  plotted the different thermodynamic observables appearing in Eq. \eqref{eq:eta}, similar to Fig. \ref{fig:EfficiencyComponents} that we had plotted for jump protocol. Although the qualitative and quantitative behaviors of almost all the plots remain very similar to those for the jump protocol, the value of $W_{\rm out}$ is substantially higher in the region $t_{\rm asy}<0.5$ (as can also be observed from a comparison between Fig. \ref{fig:EtaPower_stiffness}(b) and Fig. \ref{fig:sudden jump}(b)). This leads to the difference in the nature of curves appearing in Fig. \ref{fig:EtaPower_stiffness}(a) for $t_{\rm asy}<0.5$.

Thus, to enjoy the perks of the concatenation, one must explore the regime where $t_{\rm asy}$ is sufficiently small. The region where $t_{\rm asy}>0.5$ does not show any interesting feature, and the efficiency shows a monotonic decay (see Fig. \ref{fig:EtaPower_stiffness}).

\begin{figure}[!ht]
\centering

\begin{subfigure}{0.48\textwidth}
\centering
 \includegraphics[width=\textwidth]{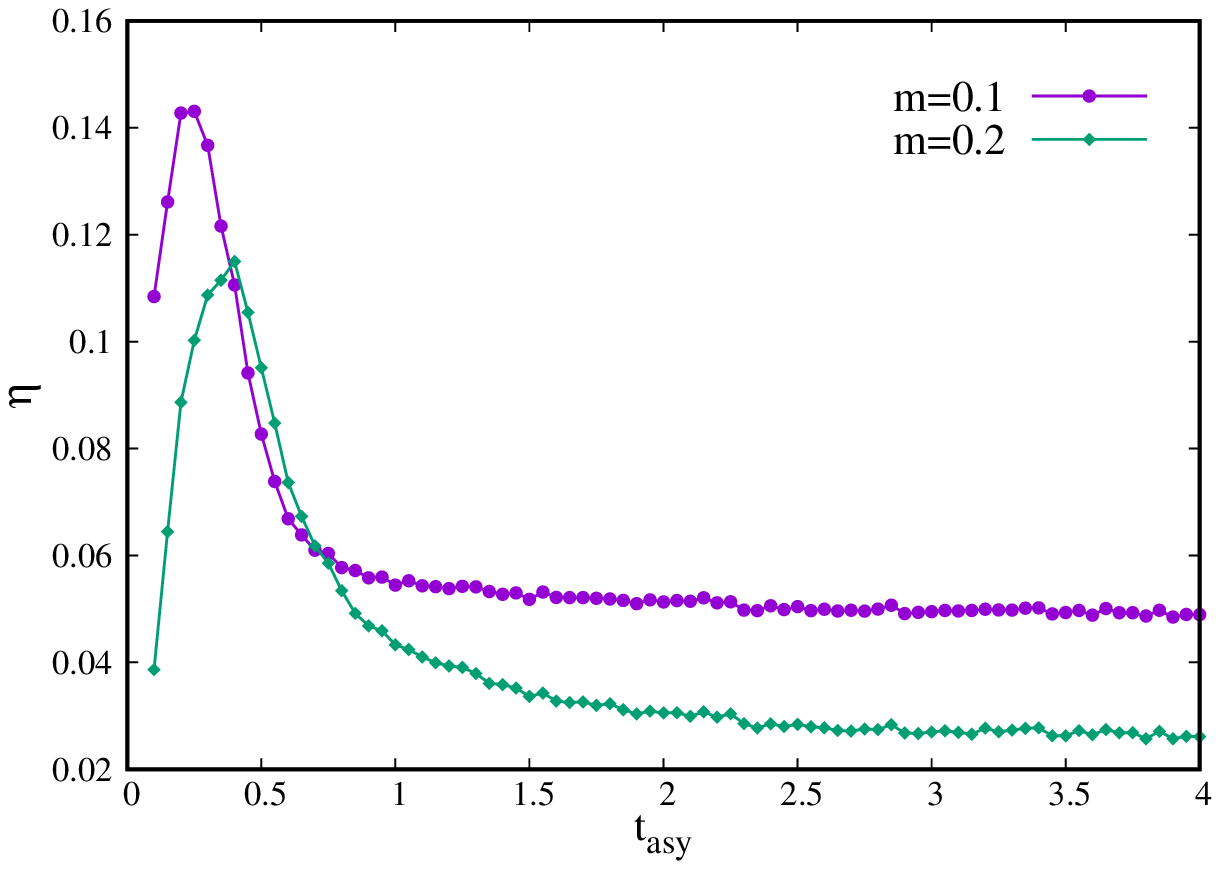}
 \caption{}
\end{subfigure}
\hfill
\begin{subfigure}{0.48\textwidth}
\centering
 \includegraphics[width=\textwidth]{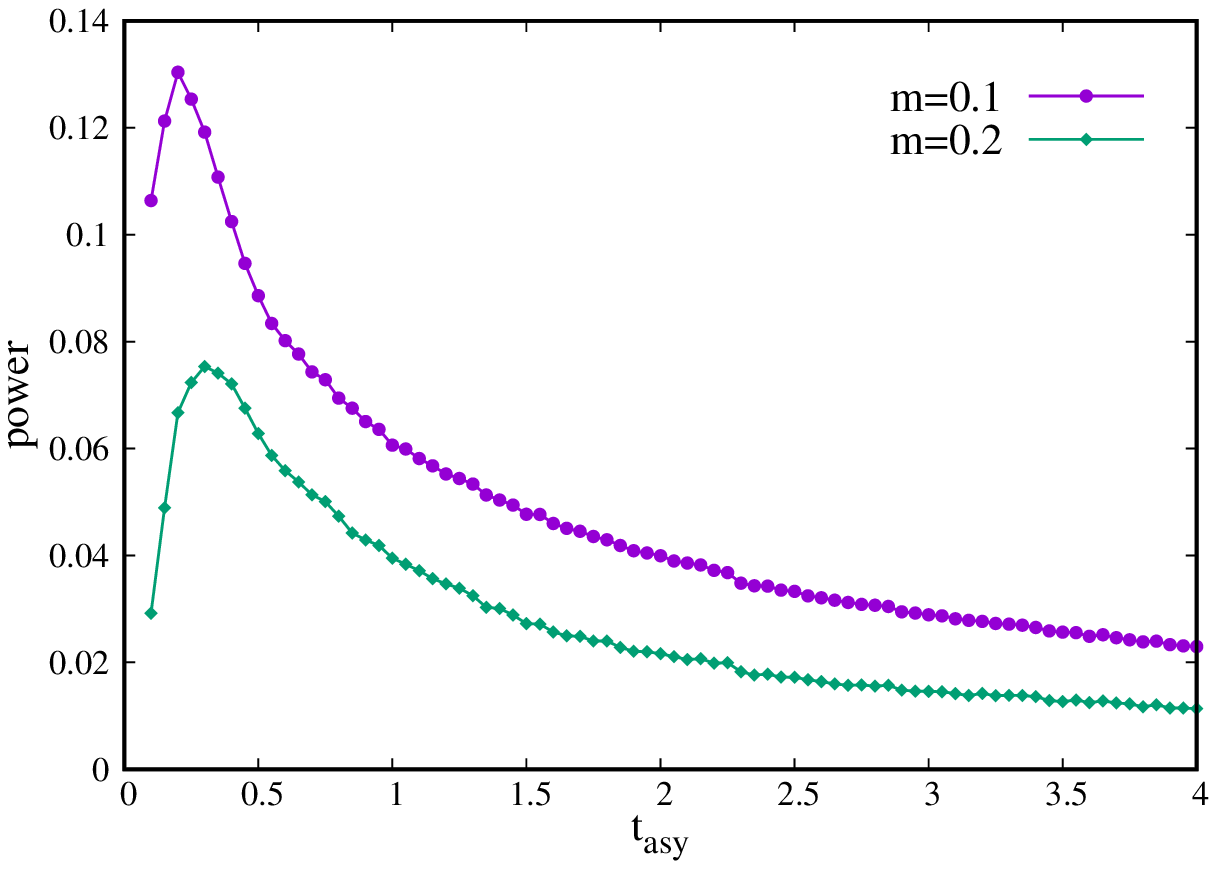}
 \caption{}
\end{subfigure}
 
 \caption{(a) Plots showing efficiency as a function of $t_{\rm asy}$ for $m=0.1$ (purple line) and $m=0.2$ (green line). (b)  Plots showing output power as a function of $t_{\rm asy}$. The parameters are: $T_1=2$, $T_2=1$, $T_3=0.1$, $k_0=10$, $\tau_2=1$.}
 \label{fig:mass}
\end{figure}

We next move to figure \ref{fig:mass}, where the variations of efficiency (figure \ref{fig:mass}(a)) and power (figure \ref{fig:mass}(b)) with $t_{\rm asy}$ for different masses has been shown. We find that if mass is increased, the values of both quantities generally decrease, while the peak value of both quantities shifts towards higher values of $t_{\rm asy}$. 
However, it is observed for a small range of the values of $t_{\rm asy}$ (approximately in the range 0.4 to 0.7), the efficiency for $m=0.2$ is slightly higher than that of $m=0.1$.

\begin{figure}[h!]
\centering

\begin{subfigure}{0.48\textwidth}
\centering
 \includegraphics[width=\textwidth]{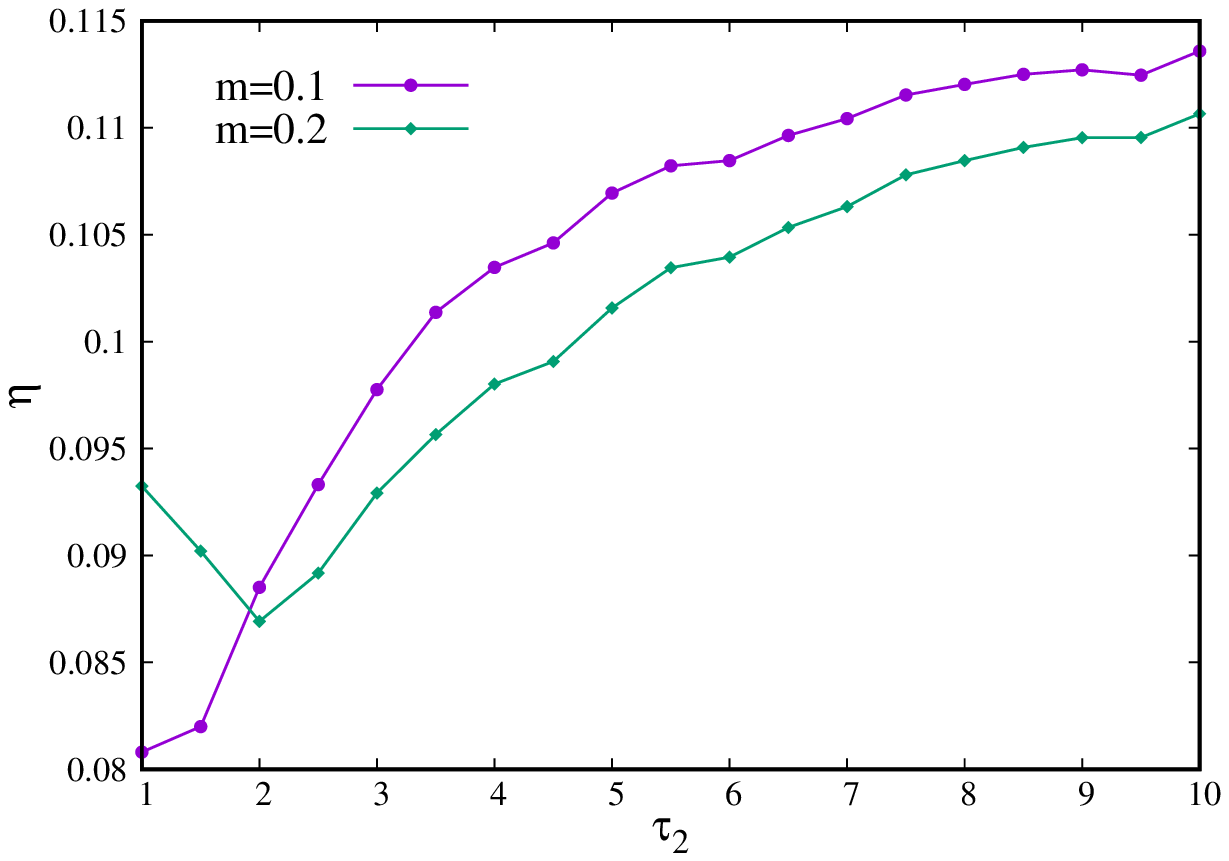}
 \caption{}
\end{subfigure}
\hfill
\begin{subfigure}{0.48\textwidth}
\centering
 \includegraphics[width=\textwidth]{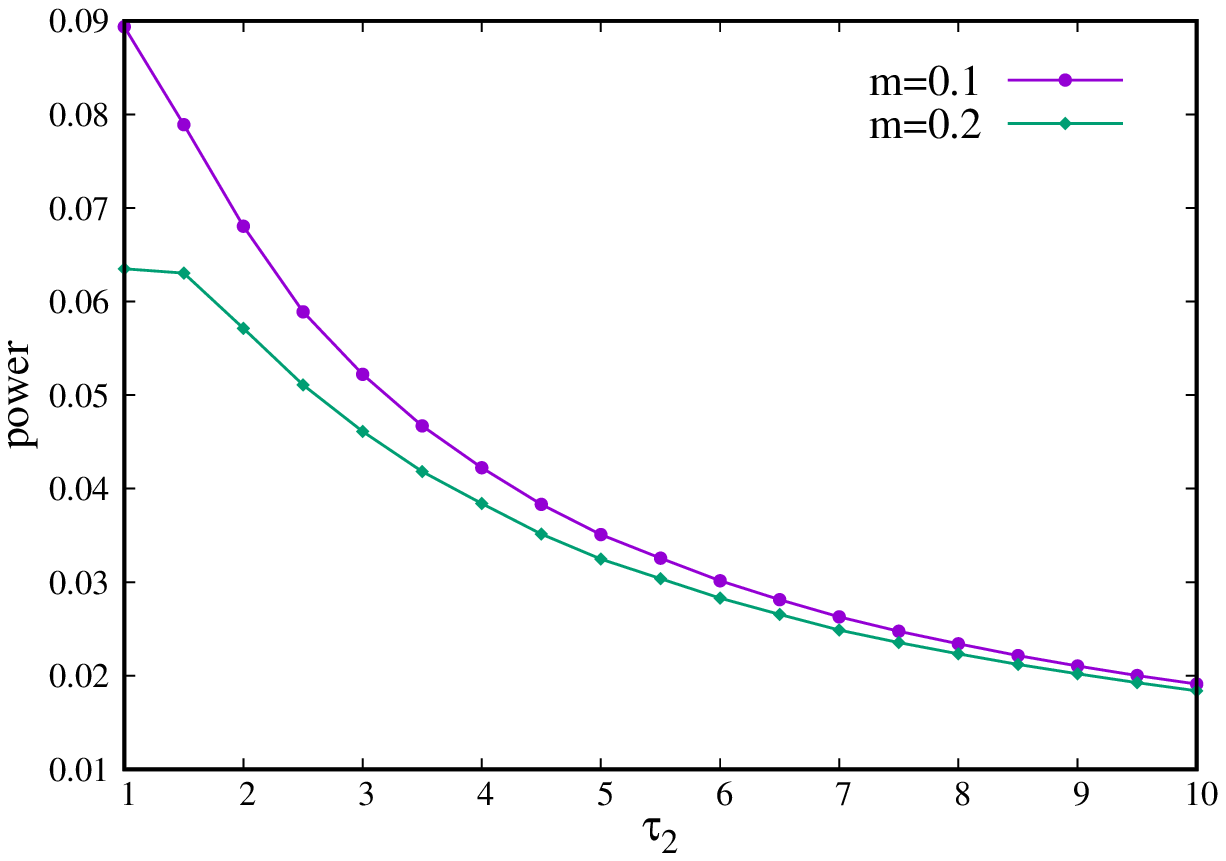}
 \caption{}
\end{subfigure}
 
 \caption{(a) Plots showing efficiency as a function of $\tau_2$, for a fixed value of $t_{\rm asy}$, for $m=0.1$ (purple line) and $m=0.2$ (green line).  (b)  Plots showing output power as a function of $\tau_2$. The parameters are: $t_{\rm asy}=0.5$, $T_1=2$, $T_2=1$, $T_3=0.1$, $k_0=10$.}
 \label{fig:tau2_eff_power}
\end{figure}

In Fig. \ref{fig:tau2_eff_power}(a), we have plotted the efficiency of the concatenated  engine as a function of the cycle time $\tau_2$ of the second engine, keeping the ratio $t_{\rm asy} = \tau_1/\tau_2$ fixed ($t_{\rm asy}=0.5$ in this case, which lies in the region of non-monotonicity). 
We find that the value of $\eta$ generally decreases for higher mass (as also observed in Fig. \ref{fig:mass}), and tends to saturate to a constant value in the quasistatic limit ($\tau_1,\tau_2 \gg 1$).
In Fig. \ref{fig:tau2_eff_power}(b), similar plots have been made for the power extracted from the engine. It is clear from the figures that both efficiency and output power depend on the absolute values of the cycle times $\tau_1$ and $\tau_2$, even if their ratios are kept constant. The plots in Fig. \ref{fig:tau2_eff_power}(b) for two different masses merge at high values of $\tau_2$, as shown in the figure, which indicates an identical power extraction (independent of mass) of the engine in the limit of slow driving.

\begin{figure}[h!]
\centering

\begin{subfigure}{0.48\textwidth}
\centering
 \includegraphics[width=\textwidth]{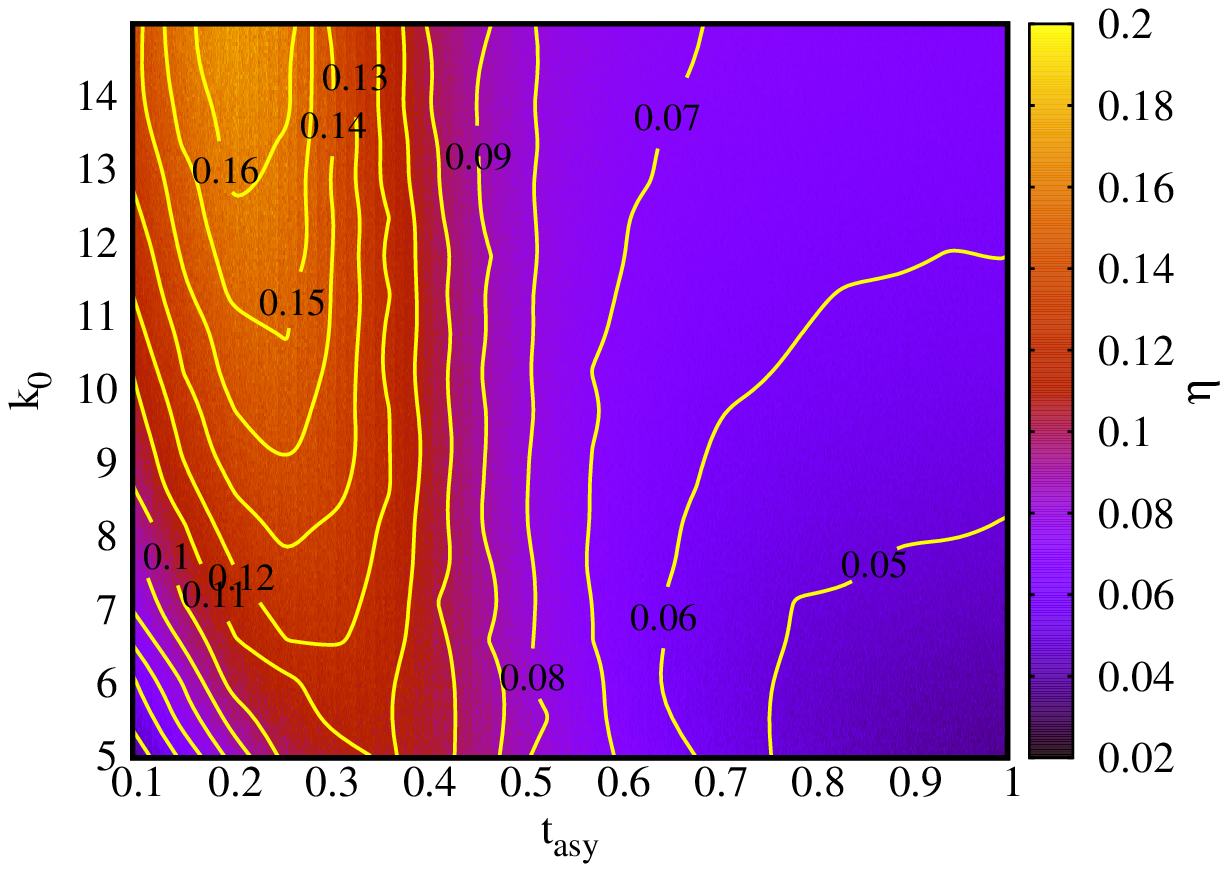}
 \caption{}
\end{subfigure}
\hfill
\begin{subfigure}{0.48\textwidth}
\centering
 \includegraphics[width=\textwidth]{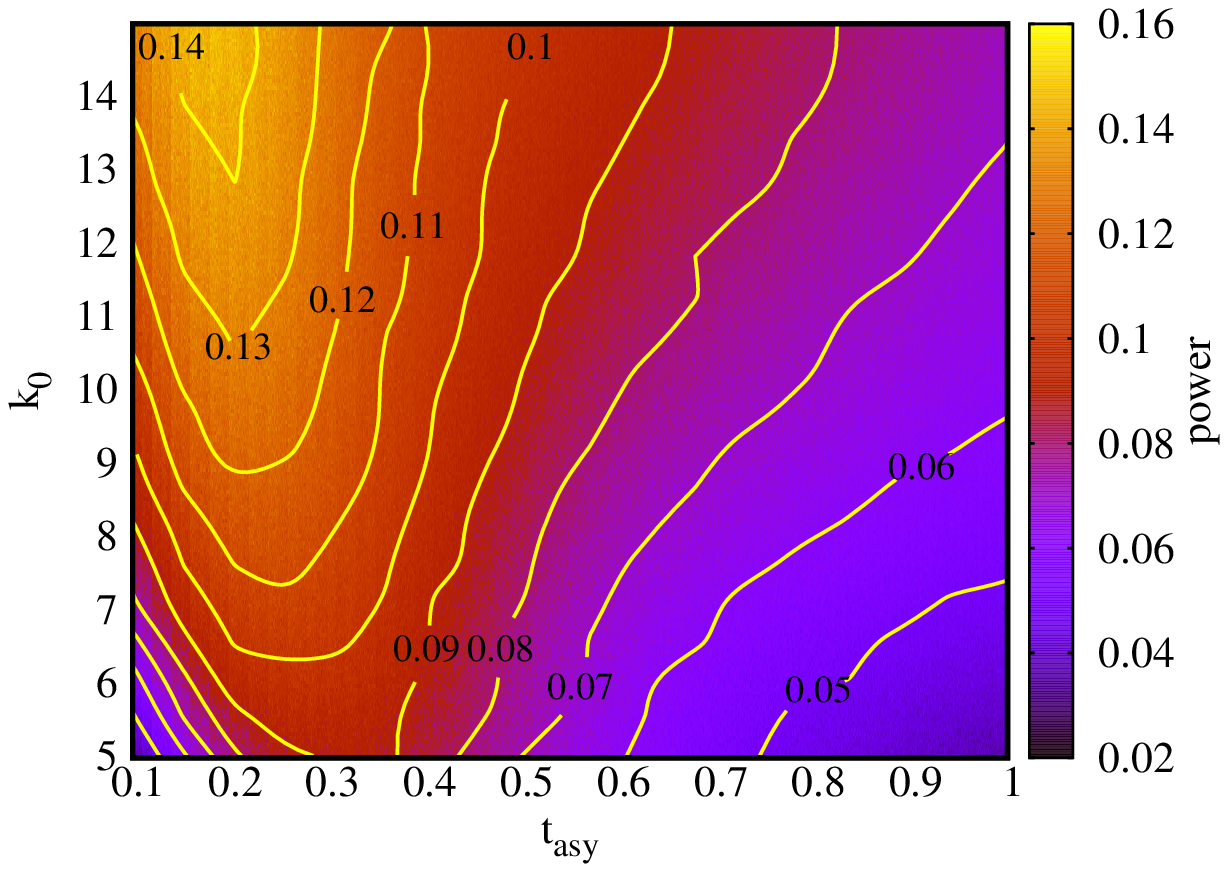}
 \caption{}
\end{subfigure}
 
 \caption{Contour plots of (a) efficiency and (b) magnitude of extracted power, as a function of $k_0$ and  $t_{\rm asy}$. The other parameters are: $T_1=2$, $T_2=1$, $T_3=0.1$, $m=0.1$.}
 \label{fig:PhasePlot}
 
\end{figure}

We study in Fig. \ref{fig:PhasePlot}(a) the dependence of efficiency on the stiffness parameter  $k_0$ and the cycle time asymmetry $t_{\rm asy}$. The non-monotonicity in efficiency as a function of $t_{\rm asy}$, visible as a tongue-shaped patch for $t_{\rm asy}\le 0.5$, in agreement with Fig. \ref{fig:EtaPower_stiffness}(a).
This non-monotonicity is also present in the functional dependence of power on $t_{\rm asy}$ as shown in Fig. \ref{fig:PhasePlot}(b), which is again consistent with the observations on Fig. \ref{fig:EtaPower_stiffness}(b). The increase of output power with the stiffness parameter is also evident.

\begin{figure}[ht]
 \includegraphics[width=0.5\textwidth]{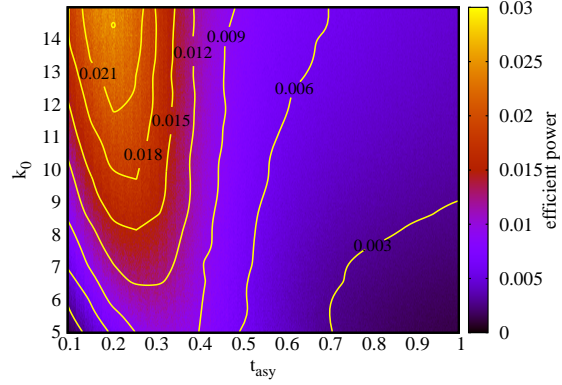}
 \centering
 \caption{Contour plot of efficient power (product of efficiency and magnitude of extracted power) as a function of  $k_0$ and  $t_{asy}$. The other parameters are: $T_1=2$, $T_2=1$, $T_3=0.1$, $m=0.1$.}
 \label{fig:EfficientPower}
\end{figure}

Fig. \ref{fig:EfficientPower} shows the dependence of the product of efficiency and power, known as \emph{efficient power} (EP), as a function of $t_\mathrm{asy}$, which is another useful quantifier for the efficacy of an engine \cite{Yilmaz2006,Johal2018}. The efficiency at maximum power (EMP) is another parameter that can be obtained from the maxima appearing in Figs. \ref{fig:EtaPower_stiffness}(a) and (b). We have separately compared the variations of EP and EMP as functions of the ratio of temperatures $T_1/T_3$, with those of the effective single engine. The latter is found to yield higher values for both parameters than the concatenated one.

Overall, the concatenation between two different nonequilibrium heat engines seems to exhibit enhanced performance mainly in terms of the output power in a suitable range of parameters. 
Both the efficiency and power show non-monotonic behavior as functions of $t_{\rm asy}$. In particular, they show peaks at much smaller values of $t_{\rm asy}$ than the symmetric case ($t_{\rm asy}=1$, i.e., $\tau_1=\tau_2$). The fact that the symmetric case can give rise to inefficient engines was also observed in the model considered in \cite{Noa_2021}.
It should thus be beneficial to identify the parameter range that is conducive to its performance. The role of the cycle time asymmetry is evident as well. The non-monotonicities of the engine outputs are observed at relatively smaller values of $t_{\rm asy}$, where the first engine is driven much faster than the second one.

\section{Three concatenated  engines} \label{sec:ThreeEngines}

\begin{figure}[h!]
 \centering
 \includegraphics[width=0.21\textwidth]{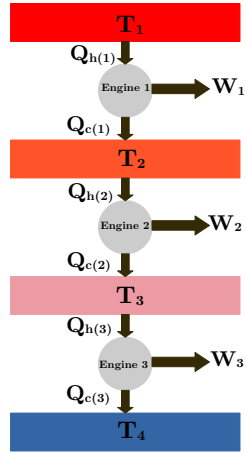}
 \caption{Schematic diagram showing three concatenated  stochastic heat engine.}
 \label{fig:3ce}
\end{figure}

We now extend the treatment for two concatenated  engines to three concatenated  engines.
 The three engines are concatenated  sequentially such that the first engine operates between two thermal baths at temperature $T_1$ (hot bath) and $T_2$ (cold bath), second engine between $T_2$ (hot bath) and $T_{3}$ (cold bath) and third engine between temperature $T_{3}$ (hot bath) and $T_4$ (cold bath). The schematic diagram is given in Fig. \ref{fig:3ce}.

\begin{figure}
\begin{subfigure}{0.48\textwidth}
\centering
 \includegraphics[width=\textwidth]{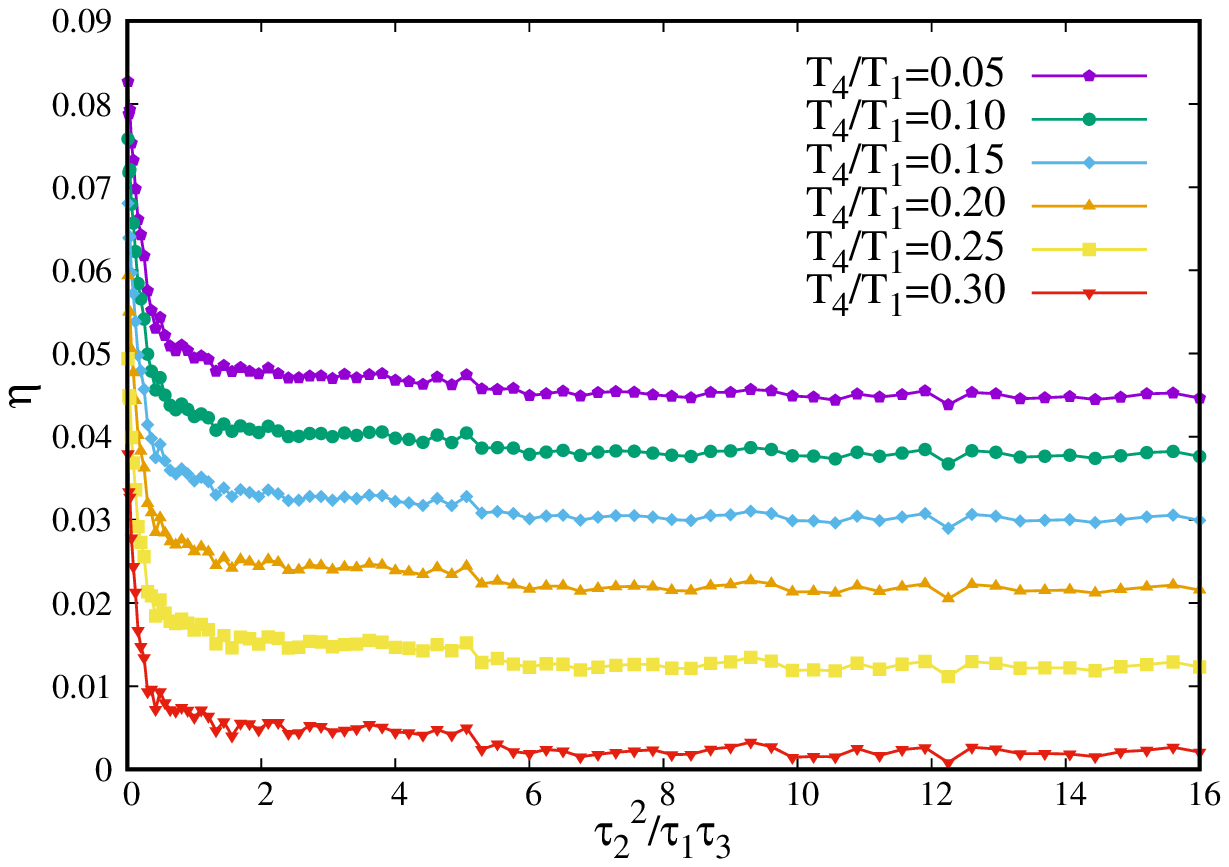}
 \caption{}

\end{subfigure}
\hfill
\begin{subfigure}{0.48\textwidth}
\centering
 \includegraphics[width=\textwidth]{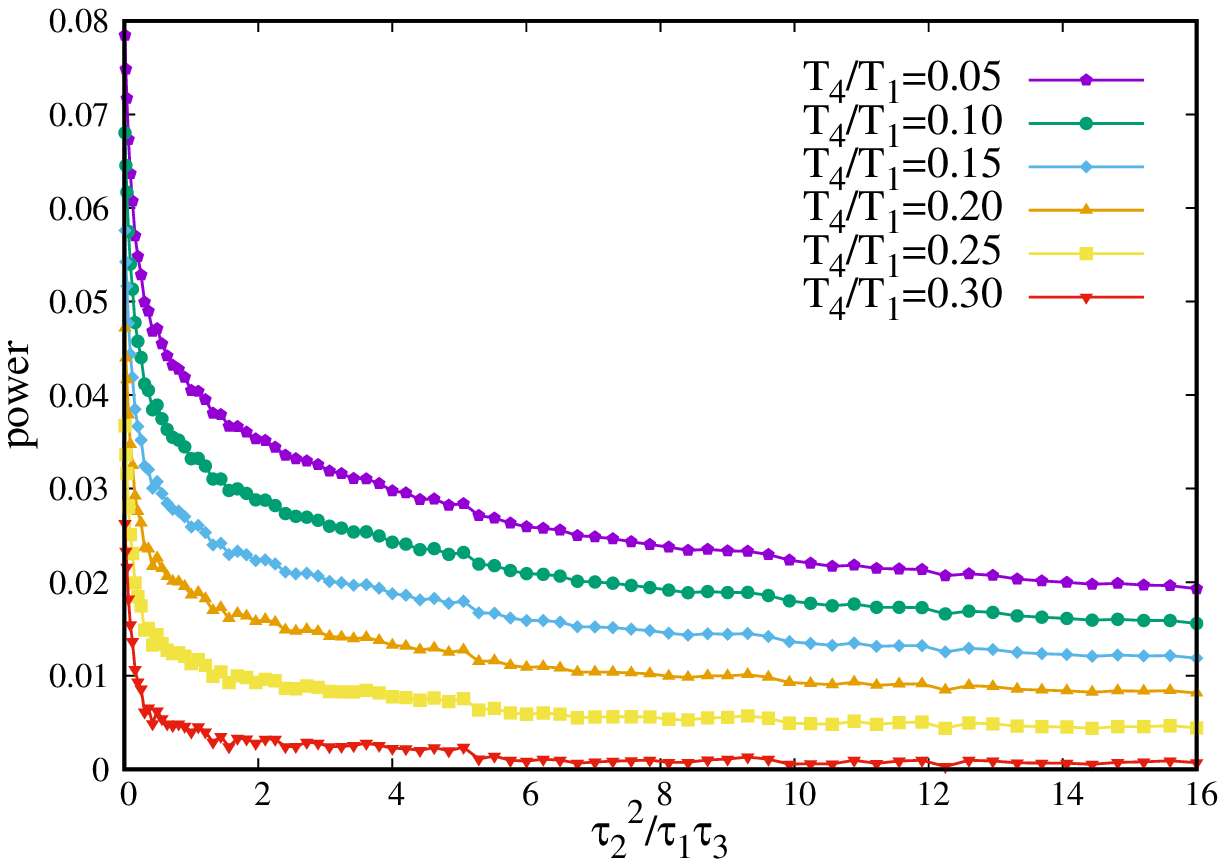}
 \caption{}
\end{subfigure}
\caption{(a) Plots showing the variation of efficiency for three concatenated  engines with $t_{\rm asy}$. (b) Similar plots for magnitude of extracted power with $t_{\rm asy}$.  The other parameters are: $T_1=2$, $T_2=1.5$, $T_3=1$, $\tau_1=1$, $\tau_3=1$, $\gamma=1$, $m=0.1$, $k_0=10$.}
\label{fig:efficiency and power 3ce}
\end{figure}

 We define the cycle time asymmetry in this case as the square of the ratio between the cycle time of engine 2 and the geometric mean of the cycle times of engines 1 and 3:
 \begin{align}
     t_{\rm asy} &= \frac{\tau_2^2}{\tau_1\tau_3}.
 \end{align}
 Fig. \ref{fig:efficiency and power 3ce} (a) and (b) show the variations in the efficiency and power as a function of $t_{\rm asy}$, respectively. Both these parameters show a monotonic decay with increase in the time asymmetry, as observed for several different values of the temperature ratio $T_4/T_1$.

\section{Analytical study of concatenated  engines driven quasistatically}
\label{sec:recursion}

\paragraph{Efficiency of Stirling Engine in quasistatic case.}
We now study the efficiency of multiple concatenated Stirling's engines, when each is driven quasistatically. The analytical expressions for these efficiencies are derived below. The expressions would help us understand the role played by the nonequilibrium nature of our driving protocols. In fact, we find that the efficiency of concatenated engine will always be less than that of a single engine in this regime. We have  observed violations of this relation in the case of two concatenated engines, where in a narrow range the concatenated engine outperformed the single engine (see Fig. \ref{fig:EtaPower_stiffness}(a)).

Below, we obtain the expressions for the efficiencies of single, two and three concatenated engines, and thereafter provide the general expression for $n$-concatenated engines, $n$ being any positive integer.

\paragraph{Single Stirling engine:}
First, we consider a single Stirling engine in the quasistatic regime where the engine works between the baths at temperature $T_1$ and $T_2$, with $T_1>T_2$. 
The \emph{total} work done in the entire cycle is given by \cite{lahiri2020}
\begin{align}
    \av{W_{1}} &=-\dfrac{T_1}{2} \ln \left[\dfrac{k_{\rm max}}{k_{\rm min}}\right] \bigg(1-\dfrac{T_2}{T_1}\bigg),
\end{align}
The subscript label includes the engine number, which is trivially 1 for a single engine.
while the work done on the engine during the expansion stroke is 
\begin{align}
    \av{W_{h(1)}} &=-\dfrac{T_1}{2} \ln \left[\dfrac{k_{\rm max}}{k_{\rm min}}\right].
\end{align}
The internal energy change during the expansion stroke is
\begin{align}
    \av{\Delta E_{h(1)}} &=\dfrac{T_1}{2}\left(1-\dfrac{T_2}{T_1}\right).
\end{align}
On applying the First Law, $Q_{h(1)}=W_{h(1)}-\Delta E_{h(1)}$ (true even for a single realization), one can obtain the heat absorbed (given by $-\av{Q_{h(1)}}$) during expansion. 

The efficiency is now readily obtained:
\begin{align}\label{eta_1e}
    \begin{split}
        \eta_{1}&= \frac{-\av{W_1}}{-\av{Q_{h(1)}}} = \dfrac{\av{W_1}}{\av{W_{h(1)}}-\av{\Delta E_{h(1)}}} \\
        &=\dfrac{-\dfrac{T_1}{2} \ln \big[\dfrac{k_{\rm max}}{k_{\rm min}}\big] \big(1-\dfrac{T_2}{T_1}\big)}{-\dfrac{T_1}{2} \ln \big[\dfrac{k_{\rm max}}{k_{\rm min}}\big] - \dfrac{T_1}{2} \big(1-\dfrac{T_2}{T_1}\big)} \\
       &=\dfrac{T_1-T_2}{T_1+\alpha_1}
    \end{split}
\end{align}
where, $\alpha_1=(T_1-T_2)\big[\ln \big(\frac{k_{\rm max}}{k_{\rm min}}\big) \big]^{-1}$.

\paragraph{Two concatenated Stirling Engines:}
The schematic diagram has been shown in Fig. \ref{fig:two_coupled_engine}, with $T_1>T_2>T_3$. We denote by $W_i$, with $i=1,2$, as the works done during the $i^{\rm th}$ engine's cycle, and the corresponding dissipated heats during expansion strokes by $Q_{h(i)}$. The works absorbed by engines  1 and 2 (negatives of the works extracted) are respectively
 \begin{align}\label{Win}
      \av{W_{1}}&=-\dfrac{T_1}{2} \ln \big[\dfrac{k_{\rm max}}{k_{\rm min}}\big] \big(1-\dfrac{T_2}{T_1}\big) \nonumber\\
      \av{W_{2}}&=-\dfrac{T_2}{2} \ln \big[\dfrac{k_{\rm max}}{k_{\rm min}}\big] \big(1-\dfrac{T_3}{T_2}\big).
      \end{align}
\begin{align}\label{Qhin}
\av{Q_{h(1)}}&=\av{W_{h(1)}} -\av{\Delta E_{h(1)}} \nonumber\\
&=-\dfrac{T_1}{2} \ln \big[\dfrac{k_{\rm max}}{k_{\rm min}}\big] - \dfrac{T_1}{2} \big(1-\dfrac{T_2}{T_1}\big)
\end{align}
Similarly,
\begin{align}
    \av{Q_{h(2)}}&=-\dfrac{T_2}{2} \ln \big[\dfrac{k_{\rm max}}{k_{\rm min}}\big] - \dfrac{T_2}{2} \big(1-\dfrac{T_3}{T_2}\big).
\end{align}

The efficiency can now be readily calculated:
\begin{align}\label{eta_2ce}
    \eta_{2}&=\dfrac{-\av{W_{2}}}{\av{W_{1}}-\av{Q_{h(1)}}-\av{Q_{h(2)}}} \nonumber\\
    &= \frac{T_2-T_3}{T_2+\alpha_2},
\end{align}
where $\alpha_2=T_2+(T_1-T_3)\big[\ln(k_{\rm max}/k_{\rm min})\big]^{-1}$.
Clearly, cycle time asymmetry does not contribute to the efficiency in this limit.

\paragraph{Three concatenated Stirling Engines:}
\begin{figure}[t]
\begin{subfigure}{0.48\textwidth}
\centering
 \includegraphics[width=0.5\textwidth,height=7.5cm]{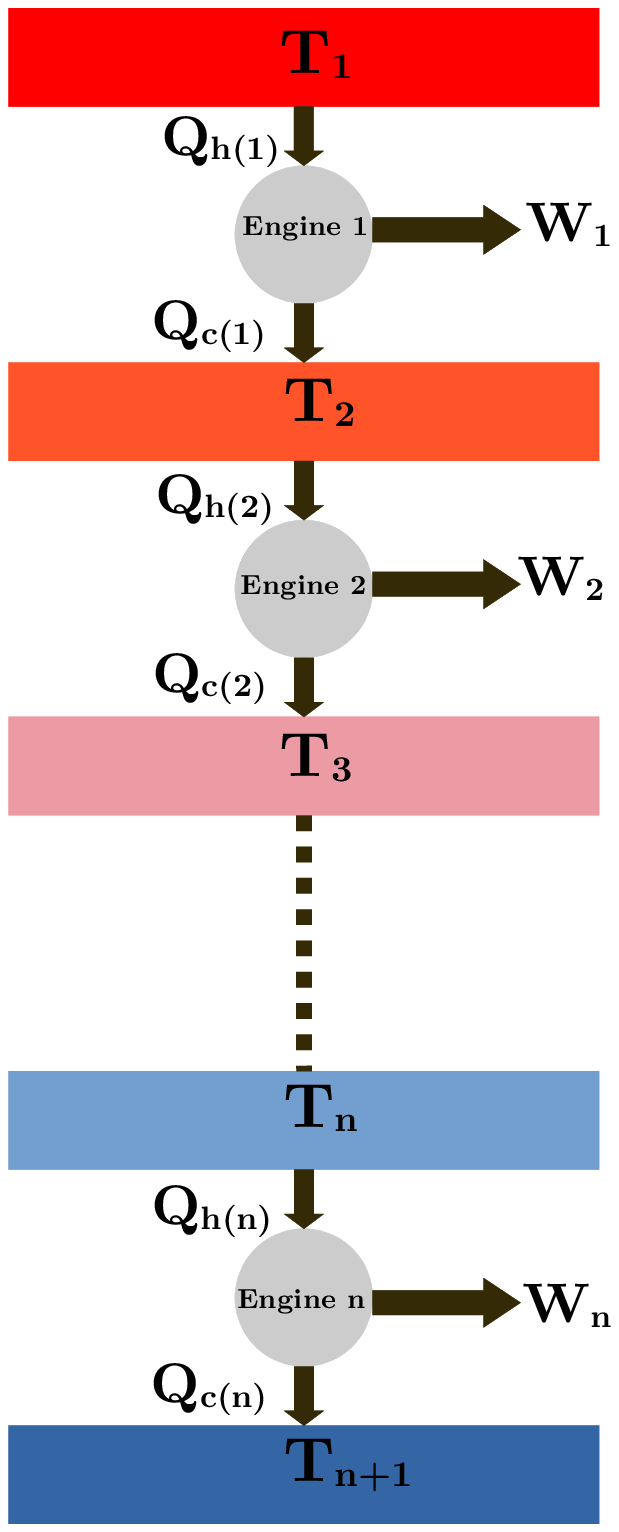}
 \caption{}

\end{subfigure}
\hfill
\begin{subfigure}{0.48\textwidth}
\centering
 \includegraphics[width=\textwidth]{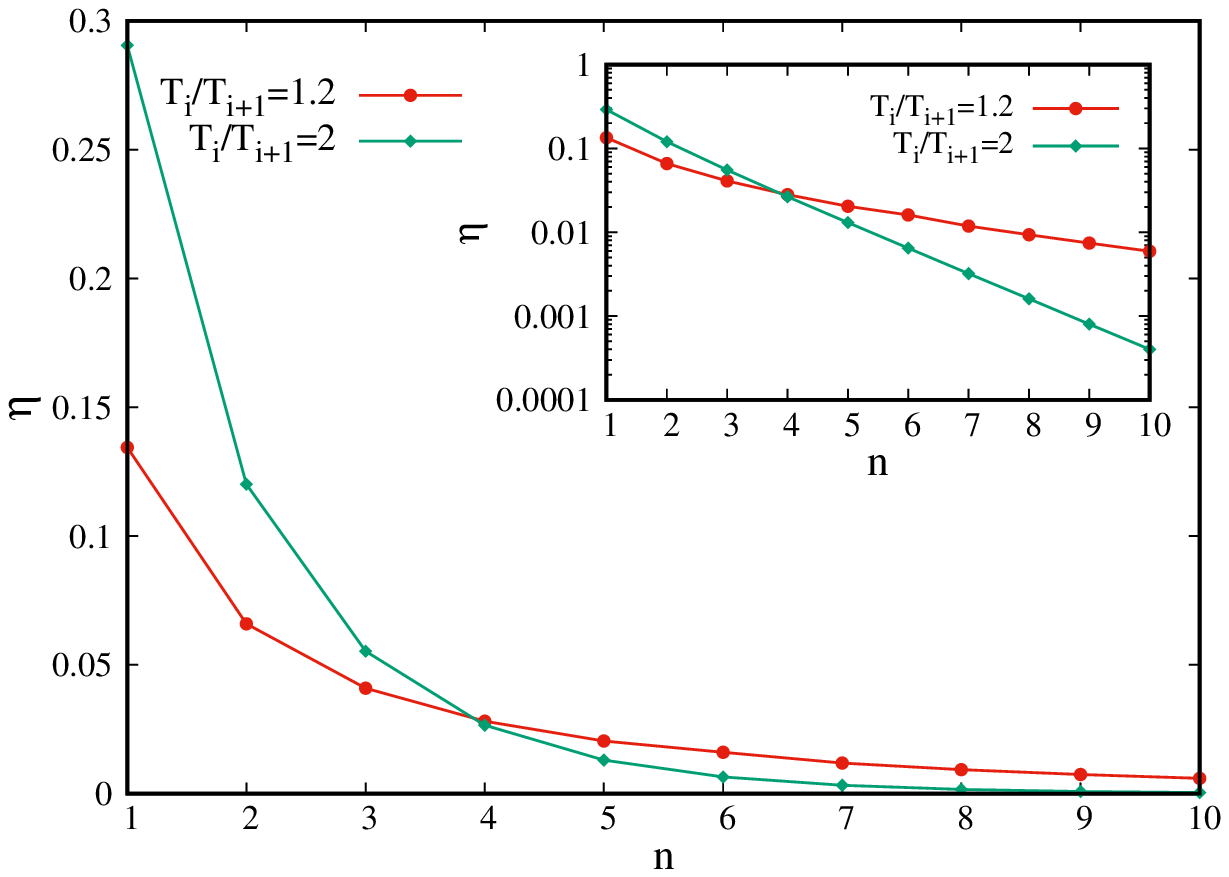}
 \caption{}
\end{subfigure}
\caption{(a) Schematic diagram showing $n$ concatenated engines. (b) Plot showing the efficiency with number of engines $n$. The ratio of the stiffness parameter is kept fixed at $k_{\rm max}/k_{\rm min}$=2. Inset shows the same plots on a semi-log scale.}
\label{fig:nce}
\end{figure}
The three engines are concatenated in a way shown in Fig. \ref{fig:3ce}, such that $T_1>T_2>T_3>T_4$. 
Proceeding as before, we obtain
\begin{align}
    \eta_{3}&=\dfrac{-\av{W_{3}}}{\av{W_{1}}+\av{W_{2}}-\av{Q_{h(1)}}-\av{Q_{h(2)}}-\av{Q_{h(3)}}} \nonumber\\
    &= \frac{T_3-T_4}{T_2+2T_3+(T_1-T_4)\Big[\ln[k_{\rm max}/k_{\rm min}]\Big]^{-1}} \nonumber\\
    &= \frac{T_3-T_4}{T_3+\alpha_3},
\end{align}
where $\alpha_3=(T_2+T_3)+(T_1-T_4)\big[\ln(k_{\rm max}/k_{\rm min})\big]^{-1}$.

\paragraph{General case: $n$ concatenated Stirling Engines:}
It is easy to extend the treatment to $n$ concatenated engines. The concatenations are as shown in Fig. \ref{fig:nce}(a), with
 $T_1>T_2>T_3>T_4>.......>T_n>T_{n+1}$. The efficiency is given by
 \begin{align}
     \eta_{n}&=\dfrac{-\av{W_n}}{\av{W_{1}}+\av{W_{2}}+\cdots+\av{W_{n-1}}-\av{Q_{h(1)}}-\av{Q_{h(2)}}-\cdots-\av{Q_{h(n)}}} \nonumber\\
     &= \frac{T_{n}-T_{n+1}}{T_n+\alpha_{n}},
     \label{eq:n_coupled}
 \end{align}
 where $\alpha_{n}=(T_2+T_3+\cdots+T_{n})+(T_1-T_{n+1})\big[\ln(k_{\rm max}/k_{\rm min})\big]^{-1}$.

The dependence on the efficiency of concatenated engines on the number ($n$) of engines concatenated is shown in Fig. \ref{fig:nce}(b), for two different values of $T_n/T_{n+1}$. Clearly, the efficiencies become smaller when $n$ becomes larger, demonstrating that coupling of engines is not very useful for quasistatic driving. The inset shows the exponential decay of $\eta_n$ with $n$. Moreover, the expression \eqref{eq:n_coupled} shows that in the limit $n\to\infty$, the numerator becomes negligibly small when compared to the denominator, so that the efficiency goes to zero:
\begin{align}
    \lim_{n\to\infty} \eta_n = 0.
\end{align}

\section{Conclusions} \label{sec:Conclusions}

Concatenated  stochastic engines have been shown to exhibit non-trivial behaviours of efficiency and power with respect to parameters like the asymmetry in the cycle times of the engines, or the magnitude of driving. We investigate the thermodynamics of such concatenated  engines in the well-known setup where each engine consists of a colloidal particle confined in a harmonic trap with time-dependent stiffness. We first take up the simple case where the stiffness is made to undergo a sudden jump in the middle of both the expansion and compression steps of each engine, where analytical solutions have been provided. The results of simulations have been compared with those obtained from numerical integration, which show a very good agreement.

The calculations become intractable when the stiffness varies linearly with time, so we have studied this case by means of simulations. In both kinds of protocols, the efficiency shows a peak when the cycle time asymmetry is small. 
In comparison to a single effective engine, we find that whereas the jump protocol does not provide any added advantage with respect to the single effective engine, the linear protocol allows one to extract a higher power than a single engine for small values of the cycle time asymmetry. 
Phase plots of efficiency and power have been plotted, each as functions of $k_0$ and $t_{\rm asy}$. They show the ranges of the latter parameters where efficiency and power exhibit non-monotonicity. 

We have also studied the variations in efficiency and power with $t_{\rm asy}$ for three concatenated  engines, but they do not show non-monotonicity. Expressions of efficiency for quasistatically driven one, two, and three concatenated  engines have been derived. Finally, a general expression for the efficiency of $n$ concatenated engines has been provided in the quasistatic limit. 
This is followed by a discussion on the general trend of the net efficiency as more and more engines are concatenated. It is shown to vanish in the limit of large $n$.

\section{Acknowledgments}

One of us (SL) thanks A. Saha for useful discussions.

\appendix

\section{Comparison of numerical results with simulations}
\label{sec:Benchmarking}

\begin{figure}[h!]
    \centering
    \begin{subfigure}{0.45\textwidth}
    \includegraphics[width=\linewidth]{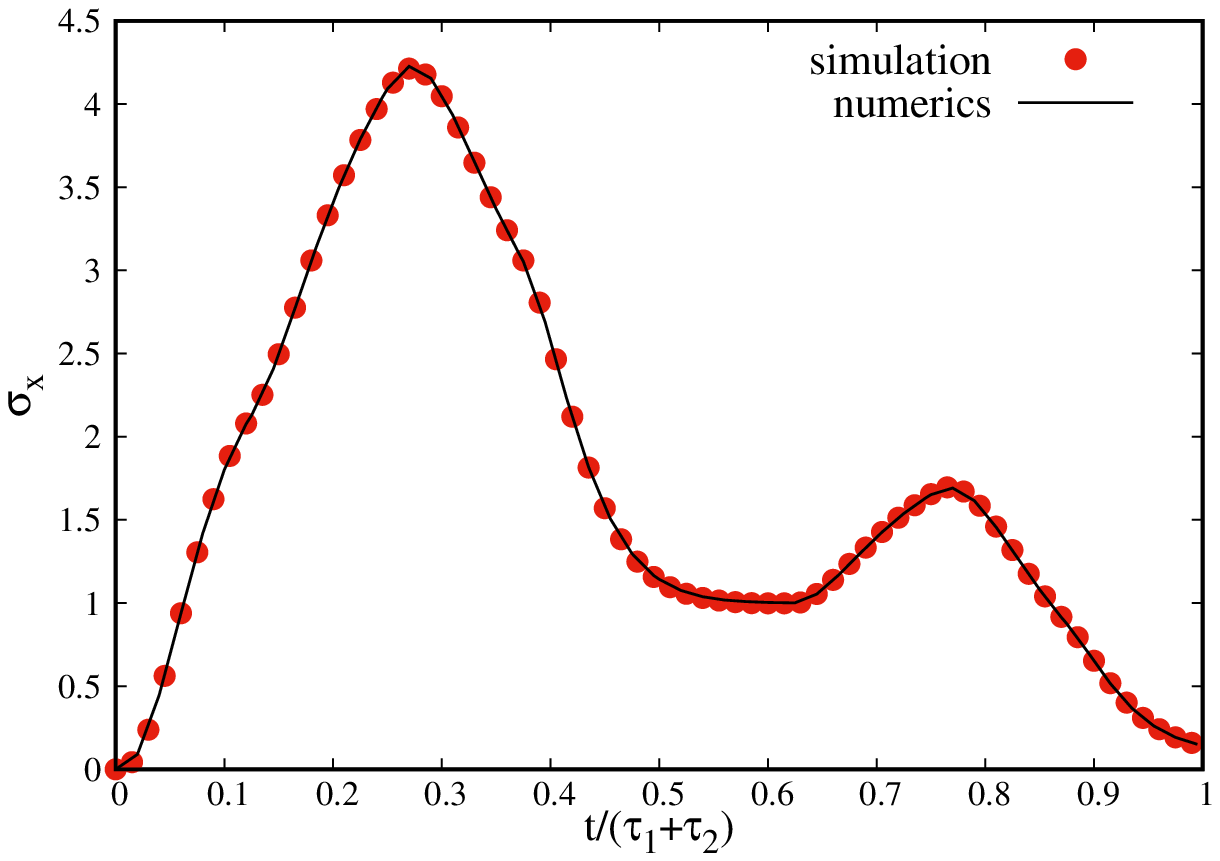}
    \caption{}
    \end{subfigure}
    \begin{subfigure}{0.45\textwidth}
    \includegraphics[width=\linewidth]{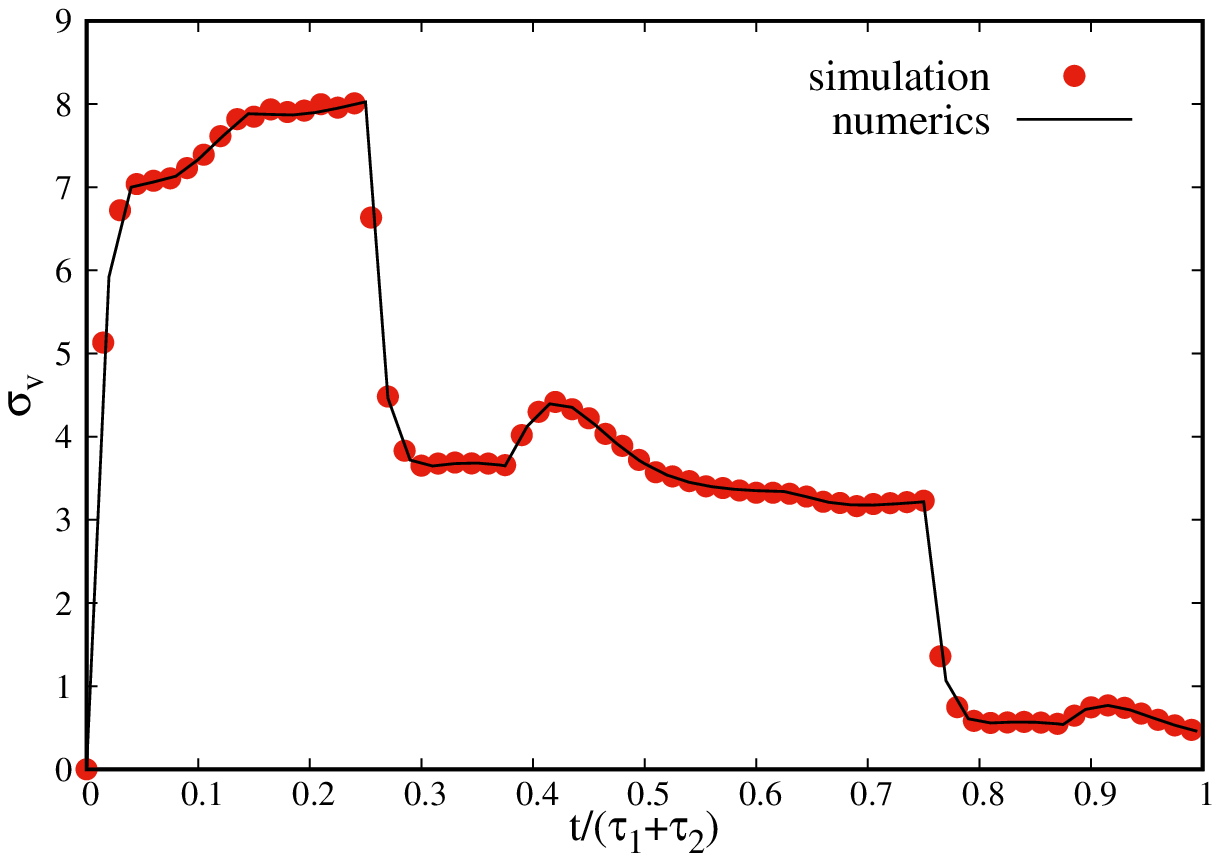}
    \caption{}
    \end{subfigure}

    \caption{(a)  Plot of the position variance obtained from numerical integration  for a full cycle of the concatenated  engine, and its comparison with simulation. Both $x$ and $v$ begin from zero at $t=0$. (b) Similar plot for variance in velocity. The parameters used in the above figures are: $m=0.3,~T_1 = 2.5,~T_2=1,~T_3=0.1, ~k_0=1, ~\gamma=1, ~\tau_1=5, ~\tau_2=5$.}
    \label{fig:Comparison}
\end{figure}

Repeating the process outlined in Eqs. \eqref{eq:ExpansionAnalytical}-\eqref{eq:AnalyticalVariances} for the second engine as well, we can obtain the variances and work in the full cycle $(\tau_1+\tau_2)$ of the concatenated  engine. 
Using these expressions we can also find analytical expressions for work and heat. However, they are too long to be included here.
Fig. \ref{fig:Comparison}(a) and (b) show the agreement of the results of numerical integration with the simulated ones, for the variances in position and velocity respectively, obtained from Eq. \eqref{eq:FirstEngine} for the first cycle of the concatenated  engine with the jump protocol given in Eq. \eqref{eq:Jump}. 
The first and the second peaks appearing in the figure correspond respectively to the transitions from the expansion step (where the variance increases with time) to the compression step (where the variance decreases with time) in engines 1 and 2.
The Heun's method of integration has been used in our simulations, and the ensemble averaging has been carried out over $10^5$ experimental realizations. 

%

\begin{figure}[!ht]
 \centering
 \includegraphics[width=0.5\textwidth]{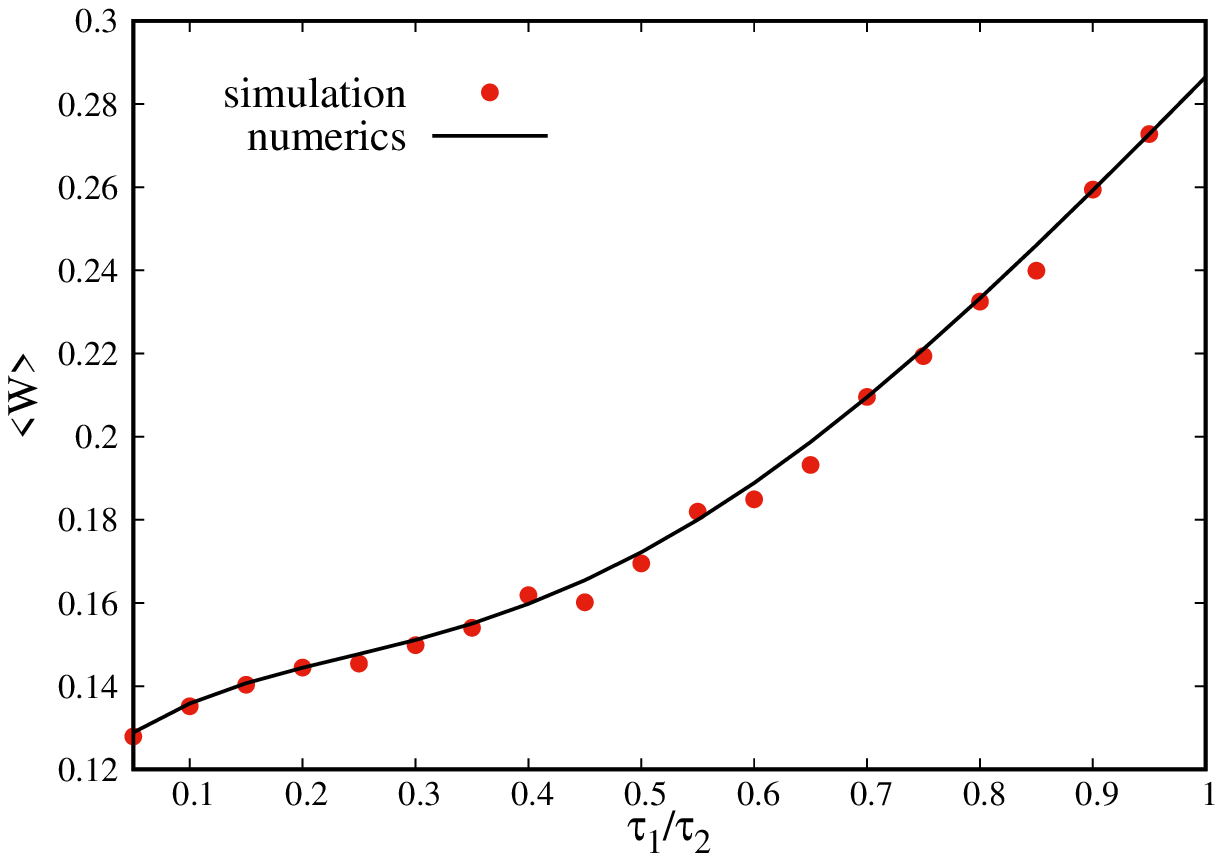}
 \caption{Plot of analytically obtained total work, i.e., $\langle W\rangle = \langle W_1\rangle+\langle W_2\rangle$, with $t_{\rm asy}$,  and its comparison with simulation. The other  parameters used in the above figures are the same as in Fig. \ref{fig:Comparison}.}
 \label{fig:W_simulation_analytics}
\end{figure}

Fig. \ref{fig:W_simulation_analytics} shows the comparison between simulation and analytics for the work done on the engine, as a function of the cycle time asymmetry $t_\mathrm{asy}=\tau_1/\tau_2$. The clear agreement between the analytical results with the simulated ones are clearly observed. 

\section*{References}

\providecommand{\newblock}{}

\end{document}